\documentclass[letterpaper]{article}

\usepackage{parskip}
\usepackage{graphicx}
\usepackage{amsmath}
\usepackage{caption}
\usepackage{chngpage}
\usepackage{amssymb}
\usepackage{textgreek}
\usepackage{graphicx}

\begin{document}
\title{“Rust Belt" Across America:\\An Application of a Nationwide, Block-Group-Level Deprivation Index
}
\author{Scott W. Hegerty\\
Department of Economics\\
Northeastern Illinois University\\
Chicago, IL 60625\\
S-Hegerty@neiu.edu
}
\date{\today}
\maketitle

\begin{abstract}

In the United States, large post-industrial cites such as Detroit are well-known for high levels of socioeconomic deprivation. But while Detroit is an exceptional case, similar levels of deprivation can still be found in other large cities, as well as in smaller towns and rural areas. This study calculates a standardized measure  for all block groups in the lower 48 states and DC, before isolating ``high-deprivation'' areas that exceed Detroit's median value. These block groups are investigated and mapped for the 83 cities with populations above 250,000, as well as at the state level and for places of all sizes. Detroit is shown to indeed be unique not only for its levels of deprivation (which are higher than 95 percent of the country), but also for the dispersion of highly-deprived block groups throughout the city. Smaller, more concentrated pockets of high deprivation can be found in nearly every large city, and some cities below 20,000 residents have an even larger share of high-deprivation areas. Cities' percentages of high-deprivation areas are posiitively related to overall poverty, population density, and the percentage of White residents, and negatively related to the share of Black residents.  

\end{abstract}

\section{Introduction}
While often thought to be confined to U.S. central cities or underdeveloped rural areas, poverty has increasingly spread to suburban areas in recent decades. This phenomenon has been detailed by Kneebone et al. (2011) and others. Understanding the spatial patterns of society's needs--and not assuming them to be confined to large central cities--is key in addressing them.

But while meeting material needs is essential, this single dimension fails to capture other important gaps. For that reason, multivariate measures of "socioeconomic deprivation" are able to include a variety of factors  in a single index. This can be used to measure a wide range of areas, both geographically and in terms of scale, and can be used to identify regions with the greatest need, so that appropriate policies can be targeted where deprivation is highest. The underlying factors behind high-deprivation areas can also be modeled empirically.

This study constructs an index of socioeconomic deprivation at the block-group level, for the entire contiguous U.S. plus the District of Columbia, keeping in mind alternative methodologies and the appropriate choice of index components. Selecting a single measure, its properties are evaluated, both for the entire country and for its largest cities. The median value for all block groups within a city is used as a measure of citywide deprivation. Then, using Detroit's value as a benchmark, other "high-deprivation" block groups are identified across the United States. These are evaluated as proportions of the total number of block groups in every state and place in the nation, and are mapped in a selection of large cities. To identify the underlying drivers of cities' proportions of high-deprivation block groups, a regression model that includes poverty and population density as controls finds that while cities with more Black or fewer White residents are likely to have larger shares of high-deprivation areas, the opposite is true for the sample of all cities. The dispersion of poverty in large cities behaves similarily to the proportion of high-deprivation areas.
	
\section{Literature}
Many previous studies of economic deprivation calculate these measures in a public health context (Carstairs, 1995), identifying covariates such as cancer mortality and life expectancy. Townsend (1987), for example, includes 77 different components in an index for Britain. In a more parsimonious measure, Salmond and Crampton (1998) use nine components for New Zealand, including housing tenure, two measures of income, employment and qualifications, adequate living space, and access to a car and a telephone. Single-parent family status is included as well. This, however makes assumptions about appropriate family status.

Besides the choice of variables, the measurement strategy is also debatable. Morris and Carstairs (1991) compare a number of measures and their relation to health outcomes, finding that some are able to explain more variation in health scores. Aaberge and Brandolini (2014) note differences in dimensions and weights. Of all possible indexing methods, Principal Component Analysis (PCA), which rotates a matrix of indivdual variables in space to extract the linear combination(s) that explain the majority of their common variance, is commonly used. Such an approach is used by Pacione (2004) in an analysis of rural Scotland, and Smith (2009) for Detroit and Portland in the United States. But a simpler, common measure is to deflate each component by its own standard deviation. This is explained below and by Hegerty (2019), who examines a nationwide sample of U.S. Census tracts, as well as the country's largest cities, and finds differences between the measures and their spatial correlations.

There is a debate about the types of areas studied as well. Bertin et al. (2014) note that many indices are normed around urban areas; a compromise index is found that applies to rural regions in France. Broadway and Jesty (1998) examine 22 Canadian inner cities, while Langlois and Kitchen (2001) apply a PCA-based index to a study of Montreal. Rural areas require their own unique components; when examining South Africa, the measure used by Noble and Wright (2013) include such variables as access to electricity. Burke and Smith (2019) include rural variables in an analysis of Norfolk, England, such as access to employment and services. While much of the literature focuses on the supposed "urban bias" of deprivation measures, Clelland and Hill (2019) examine Scotland and find little obvious bias against rural areas. 

Keeping urban and rural differences in mind, it is possible to derive an index that can be used for places of all type. This study, therefore, calculates a relatively small-scale measure of economic deprivation for the entire United States, using an appropriate choice of variables and weighting scheme. A set of large cities is examined as well. The modeling approach is described below.

\section{Methodology}
First, it is necessary to choose the appropriate variables and the weighting scheme that combines them into a single deprivation index. While there is a wide variety of potential variables, a  group of four is used here. These are based on the literature cited above and capture different types of need:
\begin{equation}
DEP_i =  \alpha PERCPOV_i + \beta PERCVAC_i + \gamma UNEMP_i + \delta NOHS_i 
\end{equation}	

All variables are taken from the U.S. Census (2015-2019 5-year American Community Survey) and cover 214,807 block groups for the 48 contiguous U.S. states as well as the District of Columbia. The poverty rate is included in order to capture material need, while the vacancy rate represents unused capacity in the physical environment. The unemployment rate represents unused labor capacity, while the percentage of residents older than 25 without a high-school diploma measures a shortage of human capital. These represent important economic capacity in consumption as well as in the main factors of production. Other possible variables that are sometimes used in the literature, which measure ethnicity or family status, are avoided here.

Next, two alternative weighting schemes are compared. In the first, PCA is used to extract the maximum common variance from the four socioeconomic variables. In the second, variances are "smoothed" by applying inverse standard deviation weights. This simpler approach makes sure that the component with the largest variance does not dominate the index. As shown below, the two are highly correlated at the national level, but since PCA is shown below to be problematic for comparing some of the smaller units analyzed here, standard-devation weights are used for the majority of this study. 

\begin{equation}
DEP_i =  \frac{1}{\sigma_1} PERCPOV_i + \frac{1}{\sigma_2}  PERCVAC_i +  \frac{1}{\sigma_3}  UNEMP_i +  \frac{1}{\sigma_4}  NOHS_i 
\end{equation}	

The national index of economic deprivation is also compared against city-level measures for the 83 cities with populations above 250,000 in 2019. For these, the median value of all block groups in the city is calculated. This is done twice--first, by simply considering each subset of values from the national index, and second, by re-calculating $\sigma$  only for the block groups within each individual city. As a third alternative, each city's value is calculated once, using Census place-level data. All three alternatives are compared against one another, and based on the results below, national weights are primarily used here.

Based on its score, as well as the city's reputation, the median value found in the city of Detroit is used as the threshold of "high" deprivation. Using weights drawn from the national sample, and calculating on citywide block-group medians, this value can be compared against other geographic areas. These include not only other cities, but also states, with high percentages of high-deprivation block groups. It is also possible to isolate individual block groups with particularly high scores. This can be shown nationally, as well as within the largest U.S. cities. As a result, it is possible to show exactly where "Detroit-level" deprivation exists, so that resources can be targeted to those specific neighborhoods.

As an additional measure of the spatial patterns behind economic deprivation, a "dispersion" score is calculated as:

\begin{equation}
Disp_i = 1- \frac{\sum A_j}{N_i}
\end{equation}
Here, $A_j$ represents "low deprivation" block groups in city \textit{i} that do not touch a high-deprivation block group, and $N_i$ is the number of low-deprivation block groups in city \textit{i}. One minus this proportion gives the share of low-deprivation block groups in a city that touch at least one high-deprivation block group. The higher the score is, the more widespread deprivation is. A low score might result if deprivation is isolated or concentrated; low-deprivation areas are in contact only with other low-deprivation areas. This dispersion is interesting in its own right--some cities have more widespread deprivation than do others--but it can also be statistically connected to other socioeconomic variables.

One major variable of interest are regions' ethnic composition. The proportion of high-deprivation areas might correspond to percentages of Black and White residents. These are plotted using bivariate methods, before being entered into a formal regression model.

In this preliminary model, places' proportions of high-deprivation block groups (\textit{\%HD}) is the dependent variable. The main control variable is the overall poverty rate; it is expected that poorer cities will have more of these areas. Population density is included as well; either sign is possible, because on the one hand, cities such as Detroit have lost population; on the other hand, crowded areas might make it hard to access resources. Finally, the percentages of Black and White residents are added in separate specifications. Additional variables may be added, based on underlying theory, but these simple specifications appear to have good explanatory power. Beacuse of the large number of places with no high-deprivation block groups, a logistic regression is first used to find out whether the explanatory variables increase the probability of having any such groups at all. Then, for the nonzero places of all sizes, as well as all 83 large cities, Ordinary Least Squares is used.

\section{Results}
\subsection{Measures and Weights}
Using both methodologies--PCA and inverse standard deviations--deprivation indices are created for 214,807 block groups in the Lower 48 states plus DC.  Figure 1 plots these measures; they are highly correlated with one another. Because of negative loadings on the principal components, the absolute values might be used to ensure a positive correlation.

\begin{figure}[ht]
        \begin{center}
\hfill
\caption{PCA vs. SD Index Values.}
\includegraphics[trim={0 .7 .7 .7in},clip,width=3in]{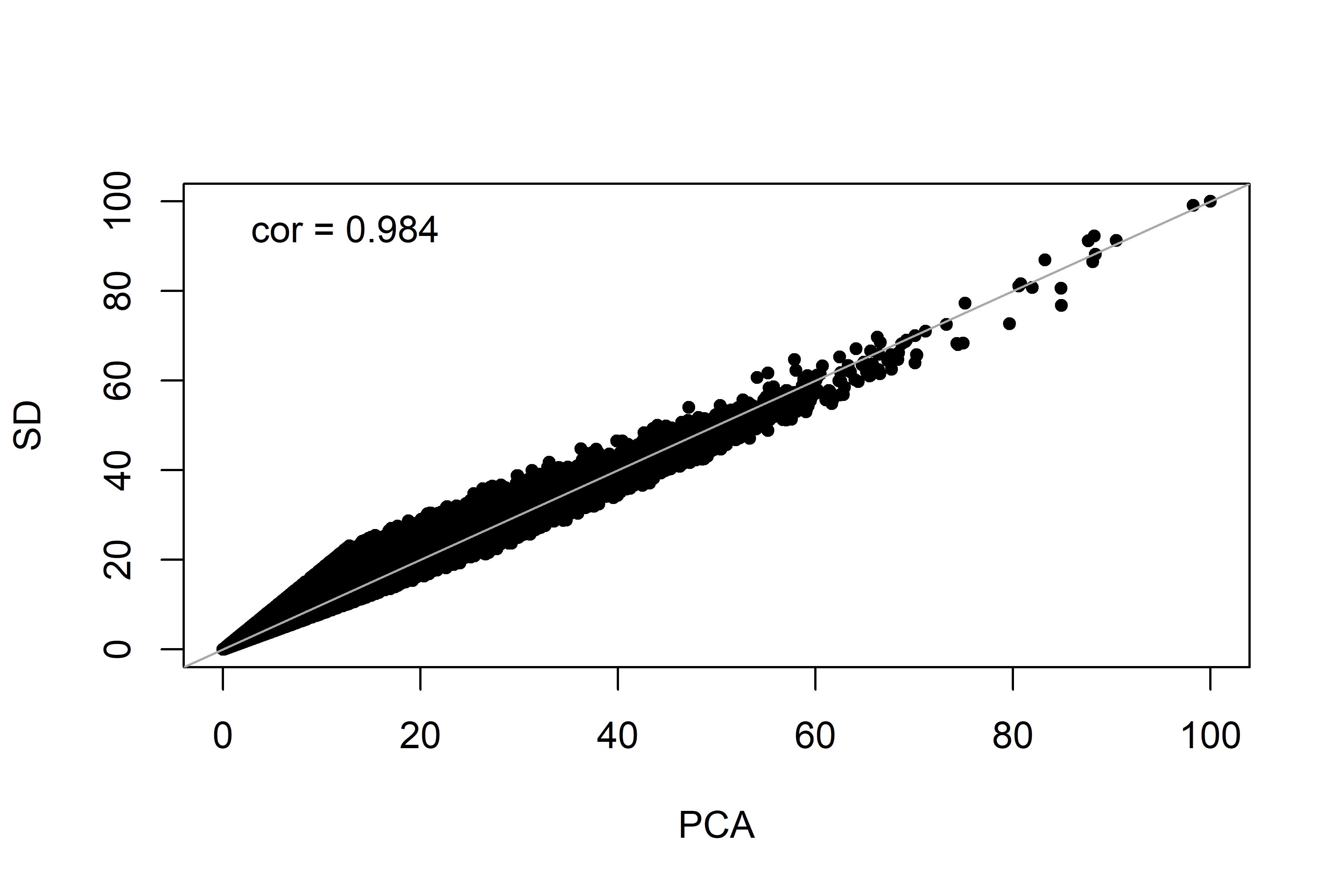}
        \end{center}
\end{figure}

These , as well as the alternative weights, are presented in Table 1. While the poverty rate has the largest factor loading (in absolute value), this variable has the smallest inverse standard deviation, and the unemployment rate has the largest.

\begin{table}[ht]
\caption{PCA Factor Loadings, Standard Deviations and Index Weights.}
        \begin{center}
\begin{tabular}{lrrrr}
&percpov&percvac&unemp&nohs\\
\hline
PCA Loading&-0.617&-0.274&-0.509&-0.534\\ 
SD&13.77&12.20&6.23&11.82\\
Weight&0.073&0.082&0.161&0.085\\
\hline
\end{tabular}
\end{center}
\caption*{}
\end{table}

When indices are calculated for individual cities, there is more difference between weights. Some cities have positive factor loadings (these are available upon request), while others have negative loadings; this makes it difficult to compare between cities. Table 2 shows the correlations between the 83 large cities' measures that are calculated three different ways using standard-deviation weights: 1) with standard deviations calculated individually for each city; 2) with identical "National" weights for al,l and 3) using place-level data rather than selecting the median value among block groups. It is clear that since the "national" measure and the place-level measure are highly correlated, that a single set of weights is appropriate to use. 

\begin{table}[h]
\caption{Correlations Between Large-City Deprivation Indices.}
        \begin{center}
\begin{tabular}{lrrr}
&City Weights&Nat'l Weights&Place\\
\hline
City Weights&1&0.51&0.42\\
Nat'l Weights&&1&0.87\\
Place&&&1\\
\hline
\end{tabular}
\end{center}
\caption*{}
\end{table}

Therefore, after creating a single, national index of socioeconomic deprivation for all U.S. block groups, city scores are created by taking the median deprivation score for all block groups within each city. Table 3 provides the summary statistics for this standard-deviation-based index, for both the country and for 83 large cities. There is a slightly higher mean score for large cities.
\begin{table}[ht]
\caption{Summary Statistics, Deprivation Index.}

  \begin{center}
\begin{tabular}{lrrrrr}
Group&Min&Mean&Med&Max&SD\\
\hline
All&0&11.73&9.94&100&7.83\\
High Pop.&0&13.89&11.97&100&9.09\\
\hline
\end{tabular}
\end{center}
\caption*{}
\end{table}

Figure 2 shows the distribution of these deprivation scores. They are right-skewed, with a large share of zero values. It is the right tail--the "high-deprivation" block groups, based on Detroit's median value of 26.413, that are of particular interest. In the entire national sample, 11,295 block groups, or 5.26\% of the total, exceed this value. Of the 39,195 block groups in the 83 largest cities, 3,855 (or 9.84\%) do so. These are close to  traditional ``right-tail'' cutoff values, suggesting that Detroit's median serves as a good threshold.

\begin{figure}[ht]
\begin{center}
\hfill
\caption{Distribution of Deprivation Scores for U.S. Block Groups.}
\includegraphics[width=.75\textwidth]{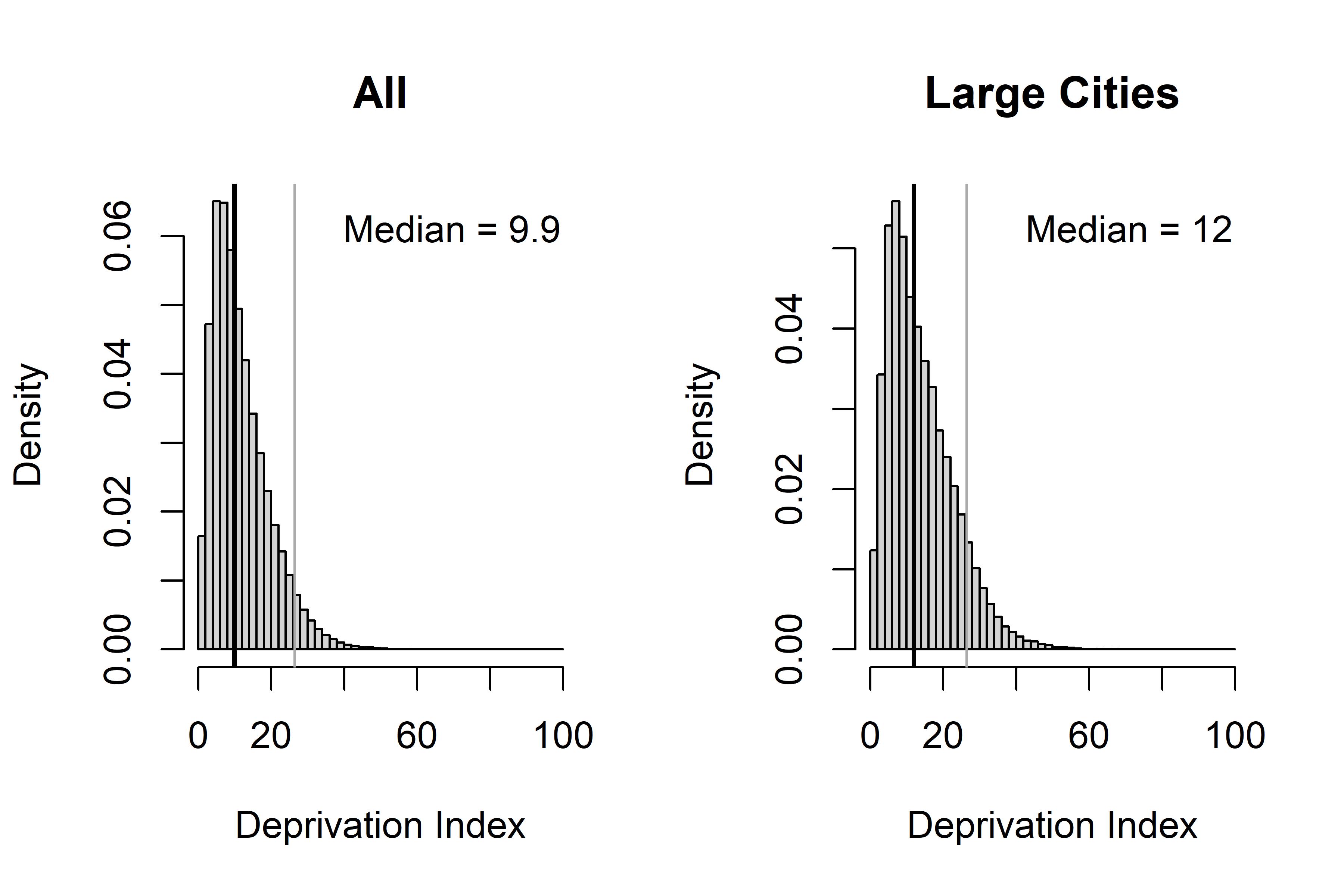}
\end{center}
\caption*{Vertical lines: Sample median and the Detroit median of 26.4 (light grey)}
\end{figure}

\subsection{High-Deprivation block groups and city characteristics}

\begin{figure}[ht]
\caption{Distribution of Socioeconomic Characteristics Within High-Deprivation Block Groups.}
  \begin{adjustwidth}{-.5in}{-.5in}  
\begin{center}
\hfill
\includegraphics[width=.6\textwidth]{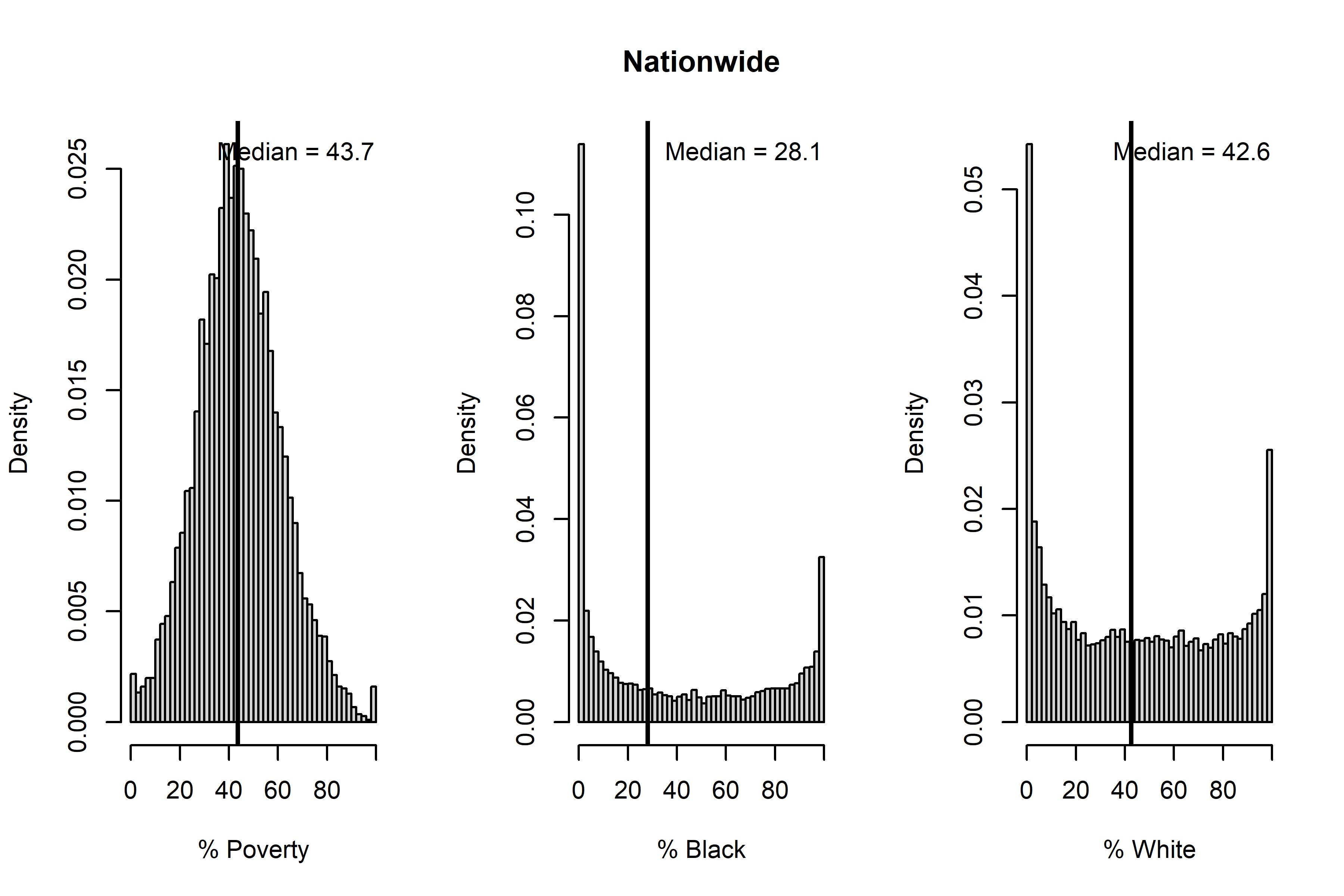}
\includegraphics[width=.6\textwidth]{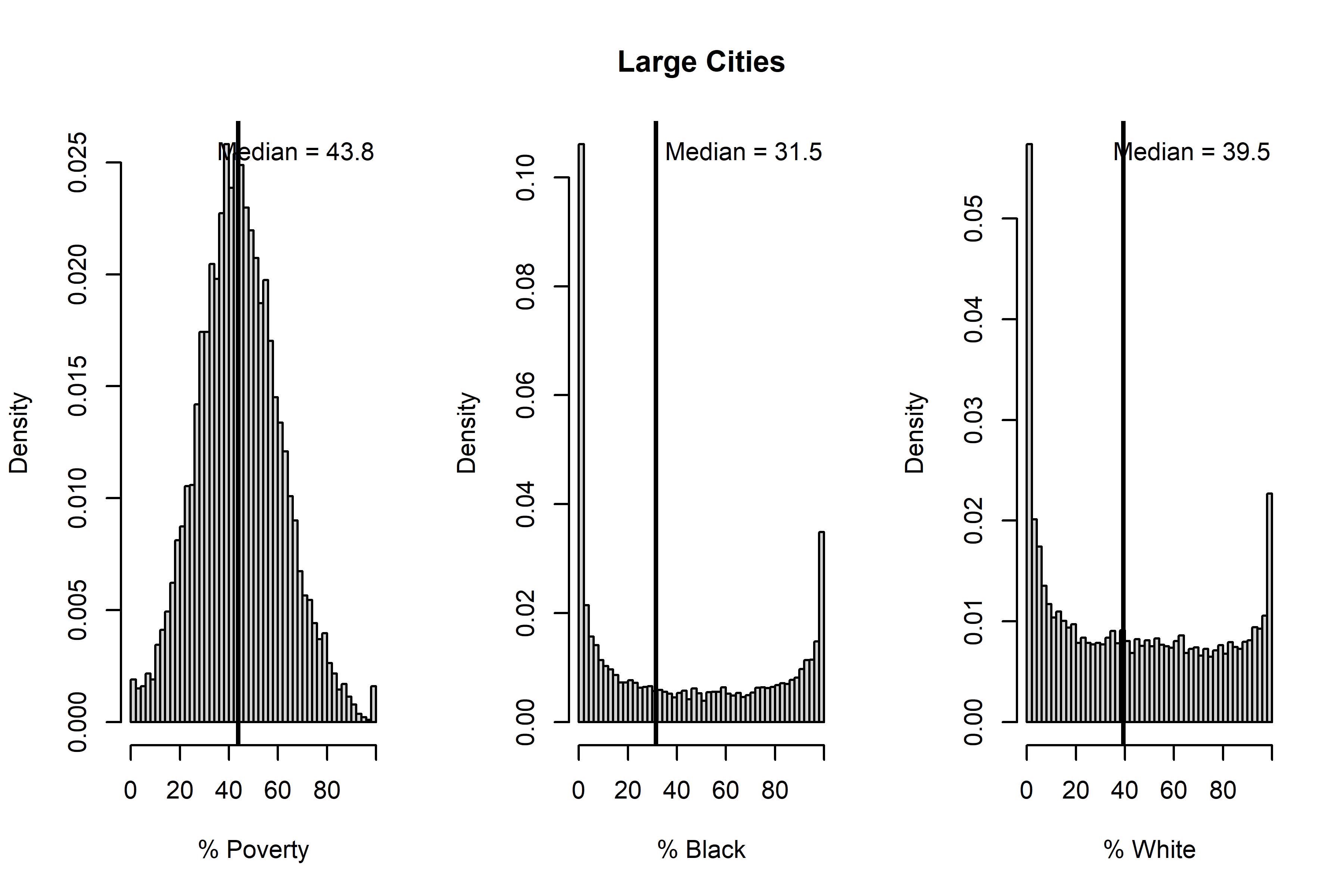}
\caption*{}
\end{center}
\end{adjustwidth}
\end{figure}

Figure 3 shows distributions for poverty rates and the percentages of Black and White residents in the high-deprivation block groups nationwide and in the set of large cities. The median poverty rate  (45.7\% vs. 45.8\%) is nearly identical in both groups, but large cities are more Black (31.5\% vs. 28.1\%) and less White (39.5\% vs. 42.6\%) on average. The distributions have similar shapes as well, with segregation shown by the relatively small share of ``intermediate'' racial percentages, compared to areas with 0\% and 100\% of a particular group.
 
Figure 4 shows a linear relationship between the percentage of high-deprivation tracts (\textit{\%HD}) and the overall deprivation score in the sample of large cities. At the same time, Detroit and Cleveland are clearly outliers, with more high-deprivation block groups relative to overall deprivation.  This suggests that these two cities comprise a unique city type, in the so-called "Rust Belt." At the same time, however, cities sich as Buffalo and Milwaukee are close to the prediction line. The first two cities, therefore, have an atypically high number of high-deprivation tracts that is not found elsewhere. A diverse group of cities that includes Anaheim, Miami, and Newark have lower-than-expected shares of high-deprivation tracts. The first two could be classified as "Sun Belt," but other cities in this category do not stand out. Likewise, Newark behaves as differently from other East Coast cities--so that it is difficult to isolate specific city types without a more rigorous investigation.

\begin{figure}[ht]
\begin{center}
\hfill
\caption{Percentage High-Deprivation (\textit{\%HD}) vs. Overall Score.}
\includegraphics[trim={0 0 0 .7in},clip,width=.6\textwidth]{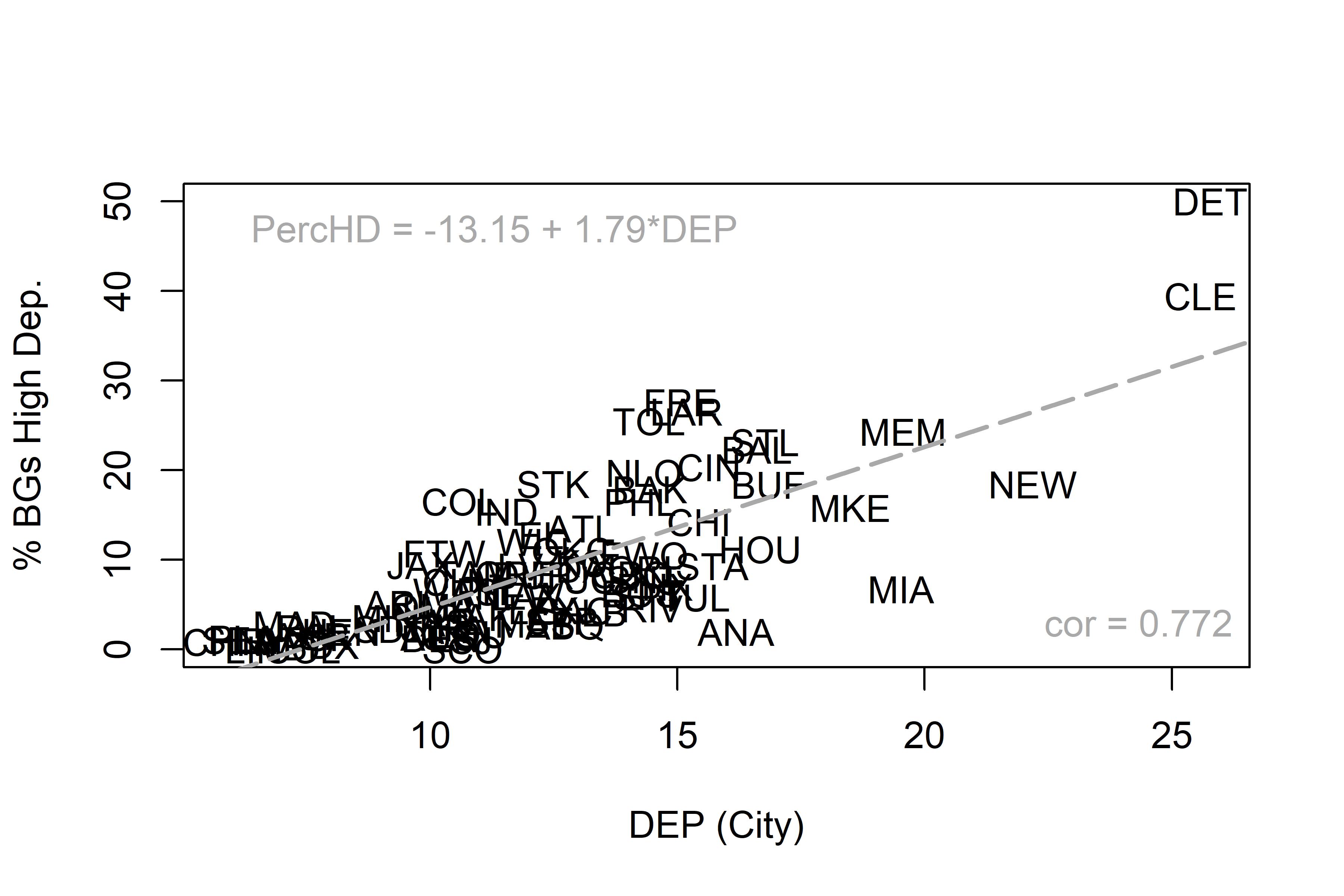}
\end{center}
\end{figure}

\begin{figure}[ht]
\begin{center}
\hfill
\caption{Large-City Percentages of High-Deprivation Block Groups\\ vs. Socioeconomic Characteristics.}
  \begin{adjustwidth}{-.5in}{-5in}  
\includegraphics[width=.6\textwidth]{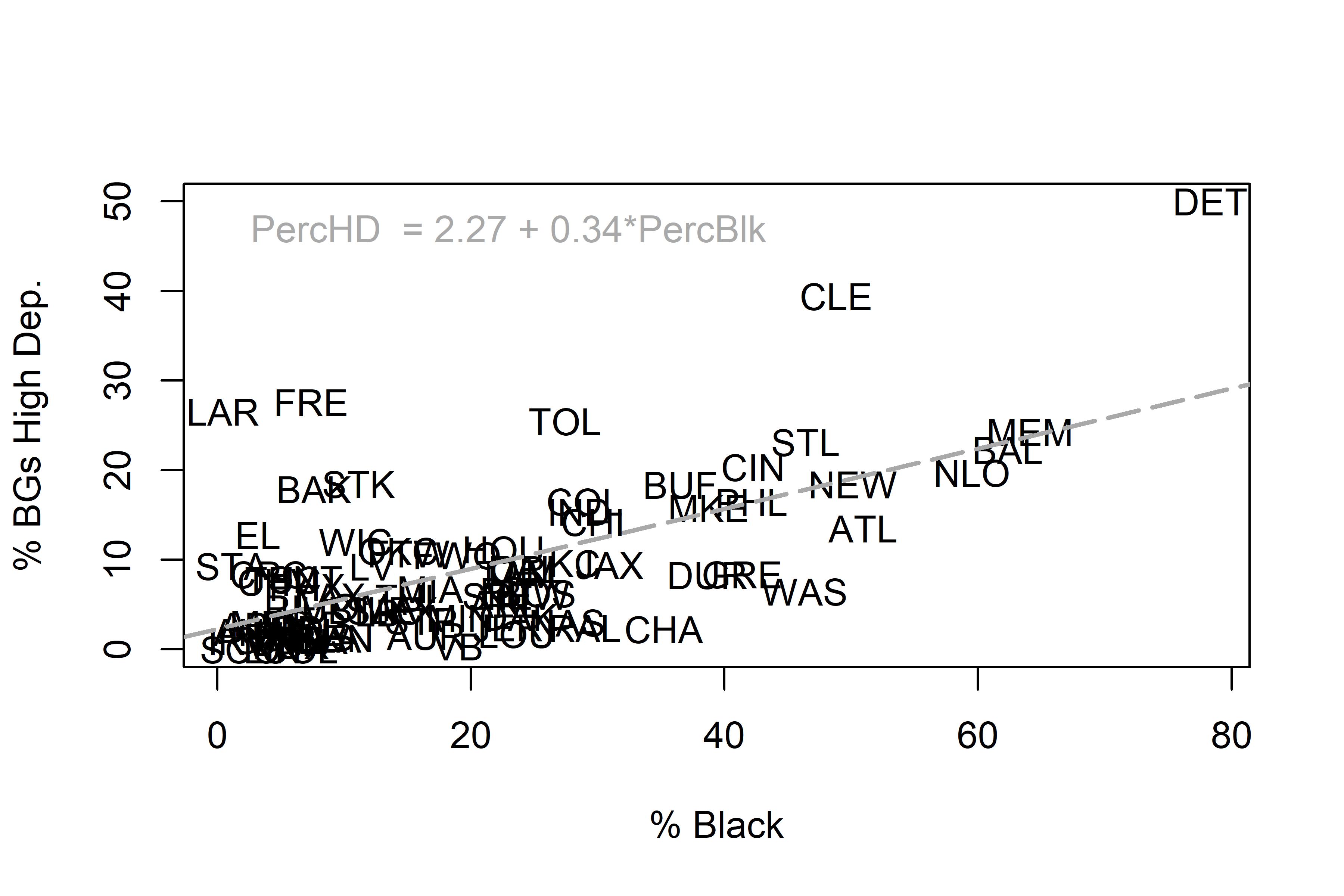}
\includegraphics[width=.6\textwidth]{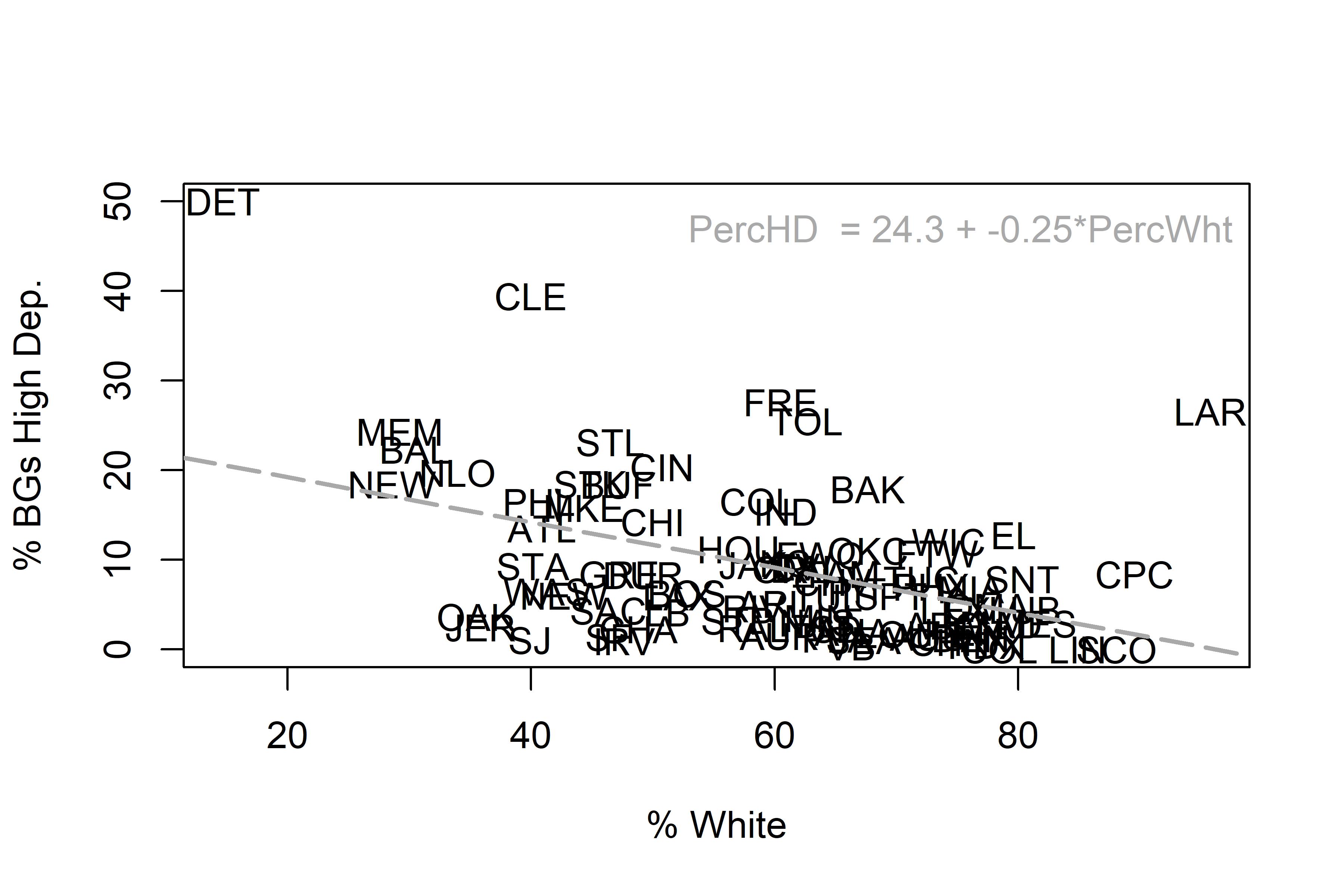}
\end{adjustwidth}
\end{center}
\end{figure}

In Figure 5, while there are clear positive and negative relationships between large cities' \textit{\%HD} and their percentages of Black and White residents, Detroit and Cleveland again stand out as outliers. As was shown to be the case above, these two cities have higher-than-predicted values of \textit{\%HD}. Washington has and lower-than-expected share relative to its share of the Black population; Atlanta does as well, to a lesser extent. Laredo's high \textit{\%HD} relative to its White population suggests that further analysis--particularly with respect to the share of Hispanic residents--is warranted. But these relationships are valid and will be added to the econometric analysis.

\subsection{Dispersion Index}
This index, which equals one if every low-deprivation block group touches a high-deprivation block group, and zero if none do, is presented in Appendix Table 9. Detroit's value of 0.872 is the highest; Virginia Beach has a score of 0.02. Of the 83 cities examined here, 16 are plotted in Appendix Figure 9. We can compare visual patterns with the calculated high-deprivation proportions and dispersion scores. Interesting pairs include Buffalo vs. Baltimore (with a similar concentration, but a higher proportion); Atlanta vs. Chicago (with a similar proportion, but a higher concentration); Madison vs. Milwaukee (with much higher scores for both proportion and dispersion); and Miami vs. Memphis. The calculated numbers capture the same urban conditions as the maps. A look at the map shows that Chicago's high-deprivation areas (on the West and South Sides) are concentrated spatially, or that Madison does not have much deprivation, but the scores calculated here allow these stylized facts to be measured empirically.

\subsection{State- and Place-Level Analysis}
Table 8 in the Appendix shows the states ranked in order of \textit{\%HD}. Mississippi has the highest, followed by Louisiana and Alabama--three states in the Deep South. These are followed by New Mexico and Michigan; again, Detroit's home state ranks high in terms of deprivation score. High deprivation in the Southwest is depicted visually in Appendix Table 10. There are correlations with poverty rates and the percentages of Black and White residents. Many of the 10 states with the lowest \textit{\%HD} values are more than 80 or 90\% white, compared with Mississippi, which is 38\% Black. 

\begin{figure}[ht]
\begin{center}
\hfill
\caption{Place-Level Percentage High Deprivation vs. Population (N = 1885)}
\includegraphics[trim={0 0 0 .5in},clip,width=.7\textwidth]{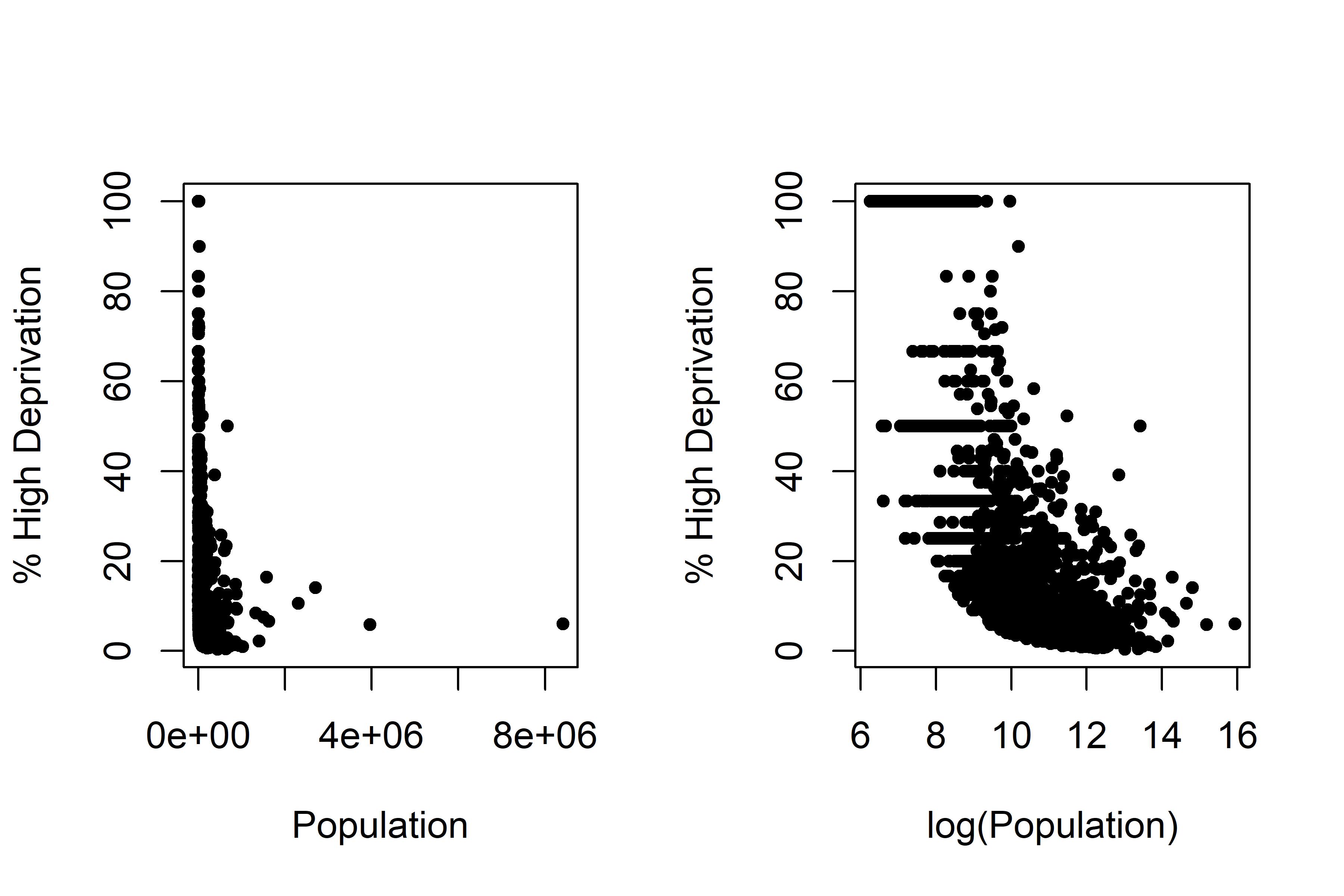}
\end{center}
\end{figure}

\begin{figure}[ht]
\begin{center}
\hfill
\caption{Distribution of High-Deprivation Percentages by Place.}
\includegraphics[width=.7\textwidth]{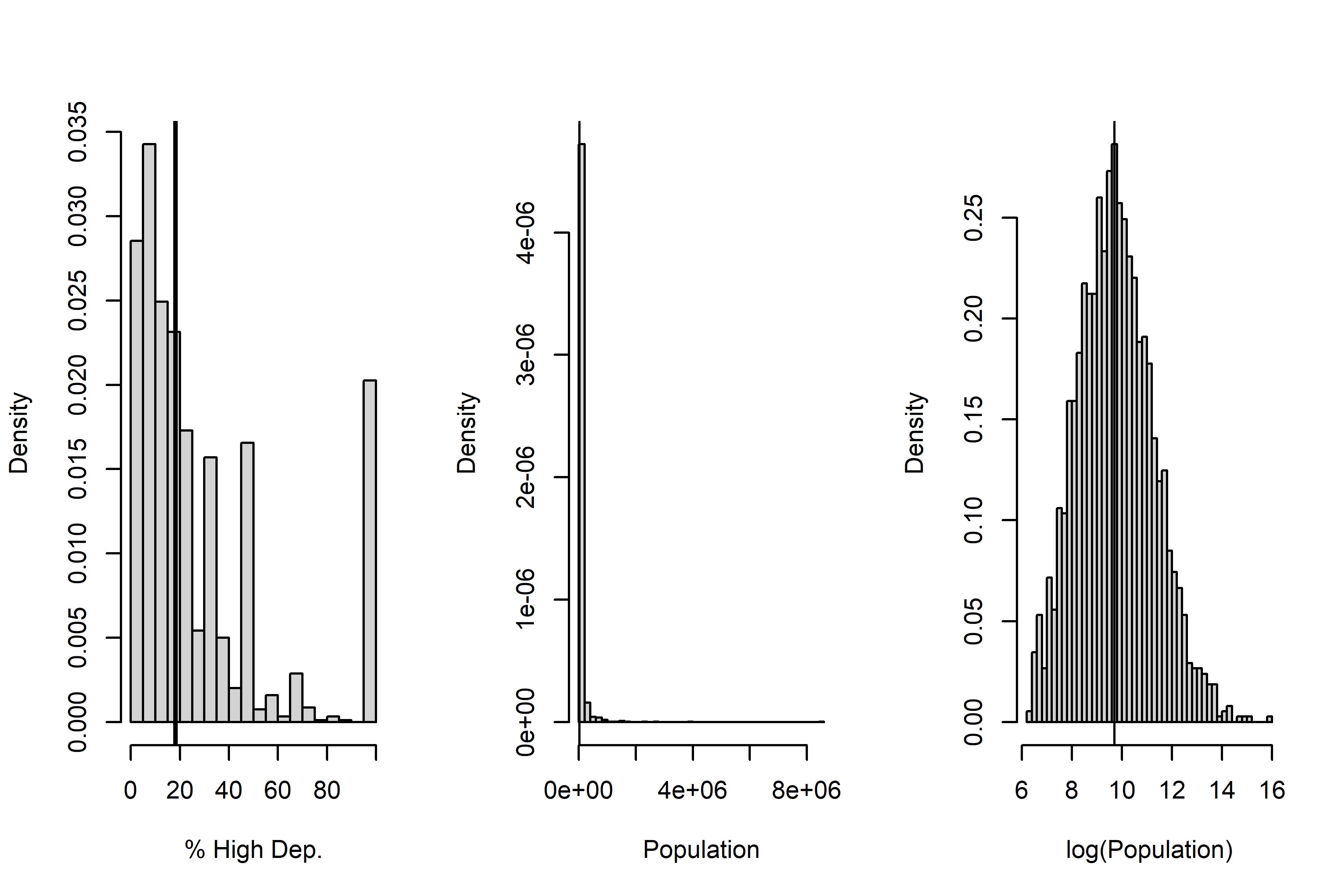}
\caption*{}
\end{center}
\end{figure}
	
This study next examines the 19,255  Census-designated places with populations above 500. Of these, 1,885 have values greater than zero, 408 Places have proportions greater than 50\%, and 191 have proportions of 100\%. Most of these are small. Figure 6 plots proportions by raw and log population; all proportions are clustered among the smallest cities.  In Figure 7, most of the 1,885 cities with nonzero values have fewer than 20\% of block groups classified as high deprivation. The median log(population) of these cities is 9.7, corresponding to a population of 16,279.

The largest cities with the highest \textit{\%HD} values are listed in Appendix Table 9; except for Detroit and Flint, Michigan, most are fairly small. Many cities with 100\% of block groups classified as "high deprivation"--which itself requires further verification to make sure this is accurate--are located in California, and almost all have populations below 10,000.

\subsection{Racial Characteristics: An Additional Look}

\begin{figure}[ht]
\begin{center}
\hfill
\caption{Median Deprivation Score by Cumulative Percentages of Racial Makeup}
  \begin{adjustwidth}{-.1in}{-.1in}  
\includegraphics[width=.5\textwidth]{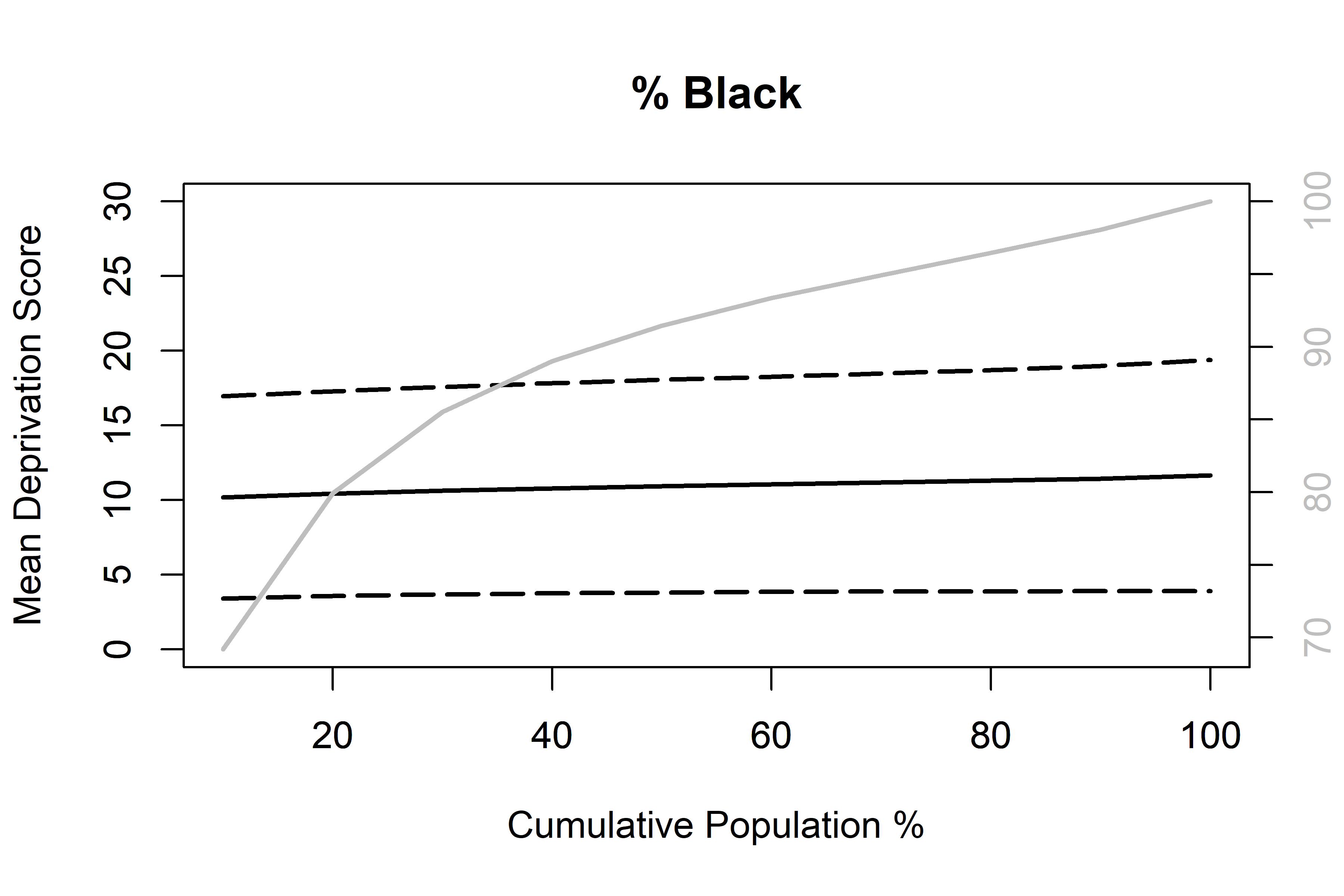}
\includegraphics[width=.5\textwidth]{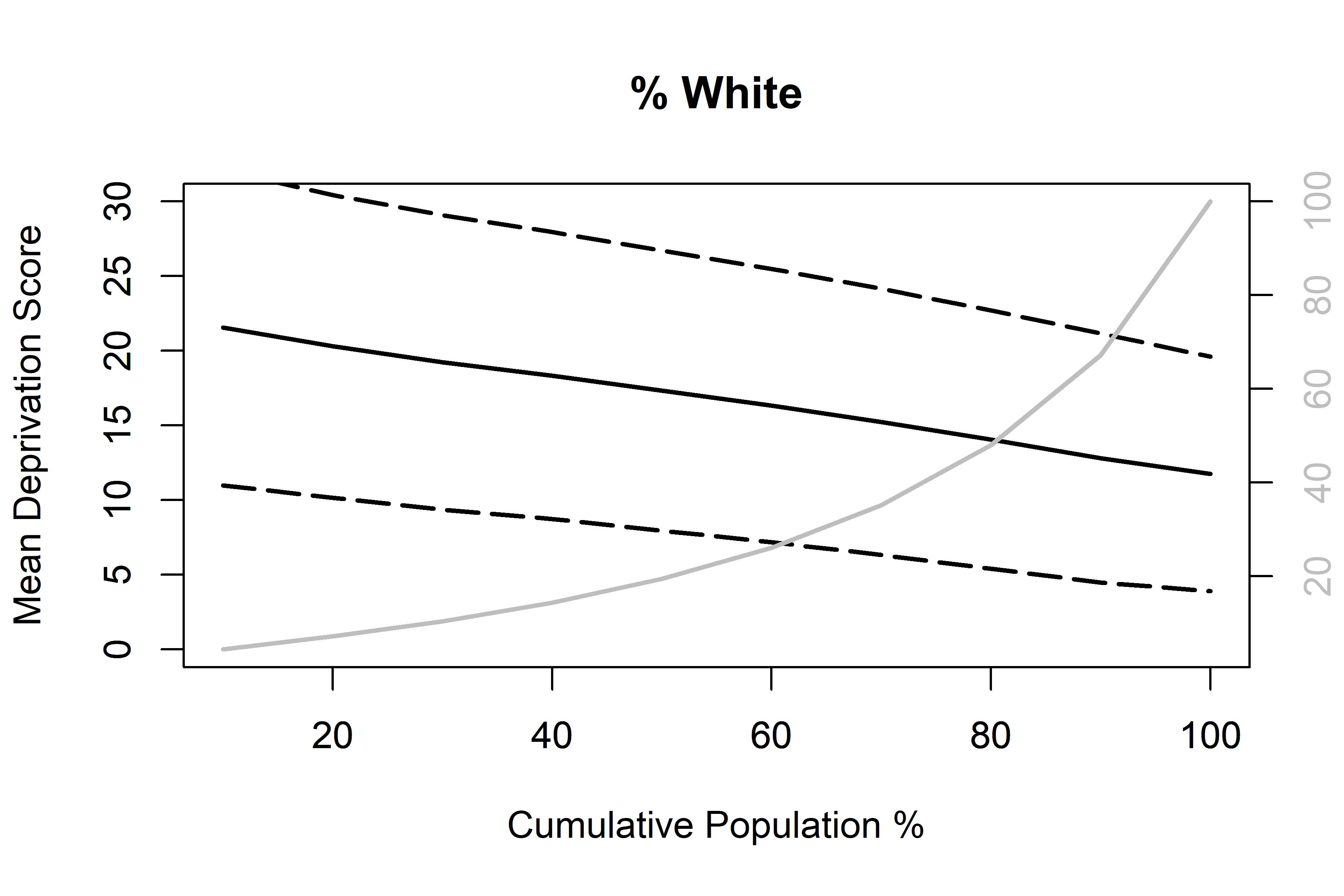}
\caption*{\footnotesize Black lines: Median score +/- 1 standard deviation from areas that are 10\% of the group population to 100\%.\\
	Grey line: Cumulative percentage of the population.}
  \end{adjustwidth}
\end{center}
\end{figure}

As one additional analysis, mean deprivation scores are calculated for increasing shares of Black and White populations, from those block groups nationwide that where a group makes up 10\% of the population, successively in 10\% increments, until the full sample (100\%) is reached. Figure 8 shows that while the average is stable across all concentrations for Black residents, it steadily declines as the sample becomes Whiter. This finding is worthy of its own investigation in more detail, but suggests that non-White (but not necessarily Black) areas tend to have higher deprivation.

\subsection{Regression Results}

These economic and racial drivers of \textit{\%HD} are entered into a regression model, to isolate the factors that influence a place's proportion of high-deprivation block groups. Because such a large share (17,370 of 19,255, or more than 90\%) have zero high-deprivation block groups, a logistic regression is first performed to find whether the explanatory variables influence the probability that a place has a non-zero proportion. These results are provided in Table 4. 

\begin{table}[h]
\begin{footnotesize}
\caption{Logistic Regression Results.}
        \begin{center}
\begin{tabular}{lrrrr}

&Coeff. (p-val.) &Coeff. (p-val.) &Coeff. (p-val.) &Coeff. (p-val.) \\
\hline
Intercept&-3.559 (0.000)&1.791 (0.000)&1.512 (0.000)&2.759 (0.000)\\
percpov&0.075 (0.000)&0.089 (0.000)&0.078 (0.000)&0.076 (0.000)\\
log(popdens)&&0.740 (0.000)&0.707 (0.000)&0.674 (0.000)\\
percblk&&&0.018 (0.000)&\\
percwht&&&&-0.016 (0.000)\\
Pseudo-$R^{2}$&0.146&0.22&0.242&0.242\\
\hline
\end{tabular}
\end{center}
\caption*{p-values in parentheses.}
\end{footnotesize}
\end{table}

Poverty rates, as expected, have a significantly positive effect on whether a place has at least one high-deprivation block group. Population density does as well, perhaps capturing that dense places have more block groups. The percentage of Black residents is positively associated with the probablility of having a nonzero proportion of high-deprivation block groups, and the percentage White is negatively associated with this odds ratio. These three variables--poverty, density, and race--explain roughly one quarter of the variance in \textit{\%HD}.

\begin{table}[h]
\begin{footnotesize}
\caption{OLS Regression Results, All Cities where \textit{\%HD} \textgreater   0.}
        \begin{center}
\begin{tabular}{lrrrr}

&Coeff. (p-val.) &Coeff. (p-val.) &Coeff. (p-val.) &Coeff. (p-val.) \\
\hline
Intercept&-8.892 (0.000)&-69.764 (0.000)&-69.257 (0.000)&-71.254 (0.000)\\
percpov&1.571 (0.000)&1.438 (0.000)&1.488 (0.000)&1.456 (0.000)\\
log(popdens)&&-8.716 (0.000)&-8.694 (0.000)&-8.649 (0.000)\\
percblk&&&-0.074 (0.003)&\\
percwht&&&&0.024 (0.348)\\
$R^{2}$&0.329&0.419&0.423&0.420\\
\hline
\end{tabular}
\end{center}
\caption*{p-values, based on robust standard errors, in parentheses.}
\end{footnotesize}
\end{table}

Next, the 1,885 places with nonzero \textit{\%HD} values are estimated using OLS. The dependent variable, which was binary in the logistic regression, is \textit{\%HD} itself. The results are presented in Table 5. One interesting finding is that, with the exception of the poverty rate, the signs on all coefficients are opposite in sign to the results in Table 4. Higher proportions are associated with lower density, lower percentages of Black residents, and higher proportions of White residents. While they require more extensive investigation, it is likely that these unexpected results are related to smaller cities in the sample. The large-city findings in Table 6 conform to expectations, as well as with the results of the logistic regression. The model with the percentage White has slightly larger explanatory power (adjusted $R^2$); the model explains almost three-fourths of the variance. We can conclude that, controlling for overall poverty, racial factors will drive the prevalence of high-deprivation neighborhoods.

\begin{table}[h]
\begin{footnotesize}
\caption{OLS Regression Results, Large Cities.}
        \begin{center}
\begin{tabular}{lrrrr}

DV = \%HD&Coeff. (p-val.) &Coeff. (p-val.) &Coeff. (p-val.) &Coeff. (p-val.) \\
\hline
Intercept&-14.248 (0.000)&-24.256 (0.000)&-22.966 (0.000)&-22.614 (0.000)\\
percpov&1.349 (0.000)&1.377 (0.000)&1.178 (0.000)&1.249 (0.000)\\
log(popdens)&&-1.471 (0.048)&-1.465 (0.045)&-2.76 (0.000)\\
percblk&&&0.108 (0.032)&\\
percwht&&&&-0.129 (0.009)\\
$R^{2}$&0.680&0.692&0.717&0.726\\
\hline
DV = Dispersion&Coeff. (p-val.) &Coeff. (p-val.) &Coeff. (p-val.) &Coeff. (p-val.) \\
\hline
Intercept&-0.185 (0.000)&-0.264 (0.009)&-0.16 (0.000)&-0.052 (0.392)\\
percpov&0.024 (0.000)&0.025 (0.000)&0.02 (0.000)&0.022 (0.000)\\
log(popdens)&&-0.012 (0.372)&&\\
percblk&&&0.002 (0.001)&\\
percwht&&&&-0.002 (0.014)\\
$R^{2}$&0.706&0.708&0.738&0.729\\
\hline
\end{tabular}
\end{center}
\caption*{p-values, based on robust standard errors, in parentheses.}
\end{footnotesize}
\end{table}

These factors also drive the dispersion of such neighborhoods within a city. In the bottom panel of Table 6, the models are repeated with the Dispersion score as the dependent variable. The results are similar, with one exception: population density plays no role. The specification with the percent Black slightly outperforms the one with the percentage White as well. Overall, therefore, we see that, controlling for overall poverty, racial makeup helps explain the concentration as well as the size of high-deprivation block groups. 
 
\section{Conclusions}
While large urban areas or remote rural regions often receive a lot of attention, socioeconomic deprivation can be found nearly anywhere in the United States. Multivariate measures of deprivation are more nuanced than simple poverty, capturing unmet needs in the physical environment, labor, and human capital. To that end, there are a number of alternative choices of component variable and weighting scheme, which need to be taken into account when examining both rural and urban areas.

This study creates an index of socioeconomic deprivation for all block groups in the contiguous United States, using four variables and variance-smoothing weights. Because of the city's unique status as a model of postindustrial depopulation, Detroit's median block-group index value is chosen as a threshold above which block groups nationwide are classifed as ``high deprivation.'' 

The results shown here confirm Detroit to be an outlier, with a median score higher than 95\% of block groups nationwide and 90\% of block groups in the country's largest cities. But large proportions of these block groups can be found in Southern states such as Mississippi, as well as in small cities in California. Such cities, many with populations below 20,000, have some of the highest proportions. Within the 83 cities with populations above 250,000, visual patterns can be compared with a calculated measure of dispersion; Detroit has the highest value, since half its block groups by definition are classified as high-deprivation. But these areas are also widespread in cities such as Cleveland, Memphis, and Newark, and more concentrated in parts of Chicago and Washington.

	A series of bivariate analyses suggest that in addition to overall poverty rates, the proportion of high-deprivation tracts within a city is closely tied to racial makeup. A logistic regression finds that an increased proportion of Black residents or a decreased proportion of White residents raises the likelihood of a U.S. Census-defined place having at least one high-deprivation block group. But for the smaller subset of places that do, the signs are reversed in an OLS regression. When the sample is restricted to the 83 large cities, the effects of race again align with the logistic regression. This suggests that small places, where deprivation is ofen extremely acute, might behave differently from large ones. Overall, Whiter cities seem to have fewer high-deprivation block groups.
	
	These results highlight the prevelance of ``Detroit-like'' deprivation nationwide, often outside of major cities and far from the Midwest or Northeast. This study isolates areas of acute need within cities, between places of all sizes, and between states. Policies could be addressed to focus on such a widespread phenomoneon, at the neighborhood, place, and the state levels. Knowing specific areas to target, as well as how this phenomenon both varies between regions yet is common nationwide, might help lead to more effective solutions.

\bibliographystyle{abbrv}
\bibliography{BG-DEP.bib}

\begin
{figure}[t]
\hfill
\caption{Deprivation and High-Deprivation Areas, Selected Large Cities.}
  \begin{adjustwidth}{-1in}{-1in}  
\includegraphics[trim={1in 0 1in 0},clip,height=1.7in]{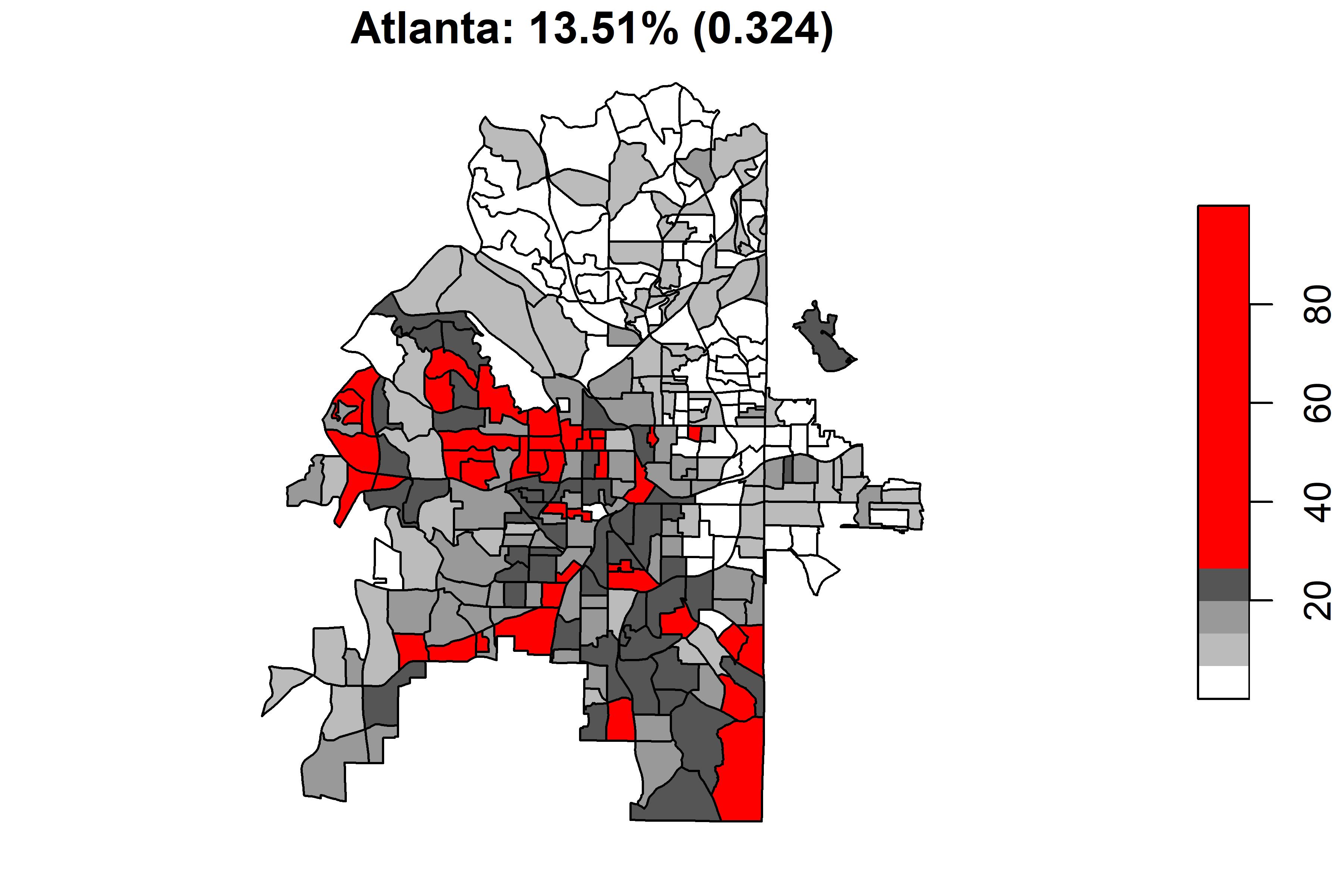}
\includegraphics[trim={1in 0 1in 0},clip,height=1.7in]{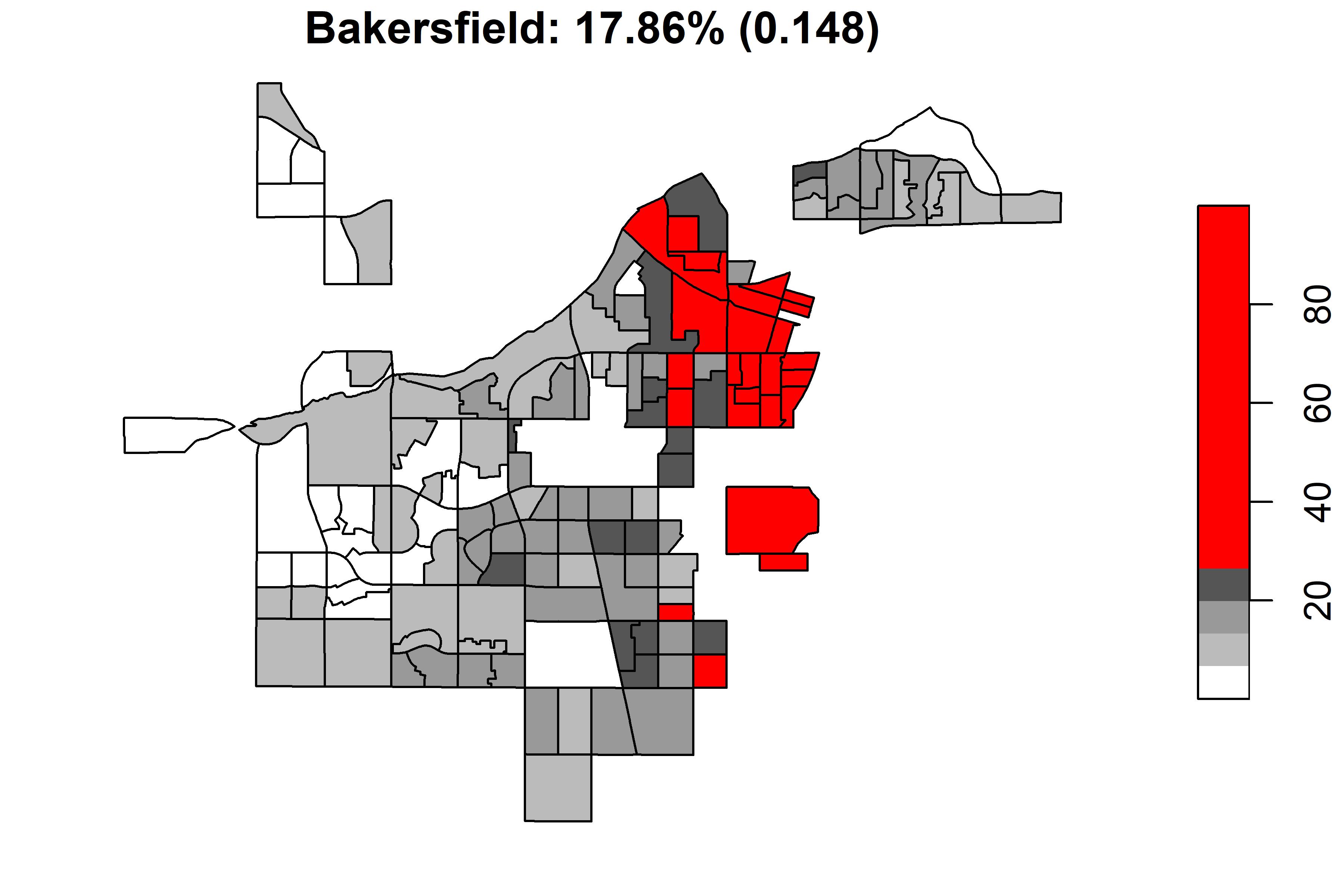}
\includegraphics[trim={1in 0 1in 0},clip,height=1.7in]{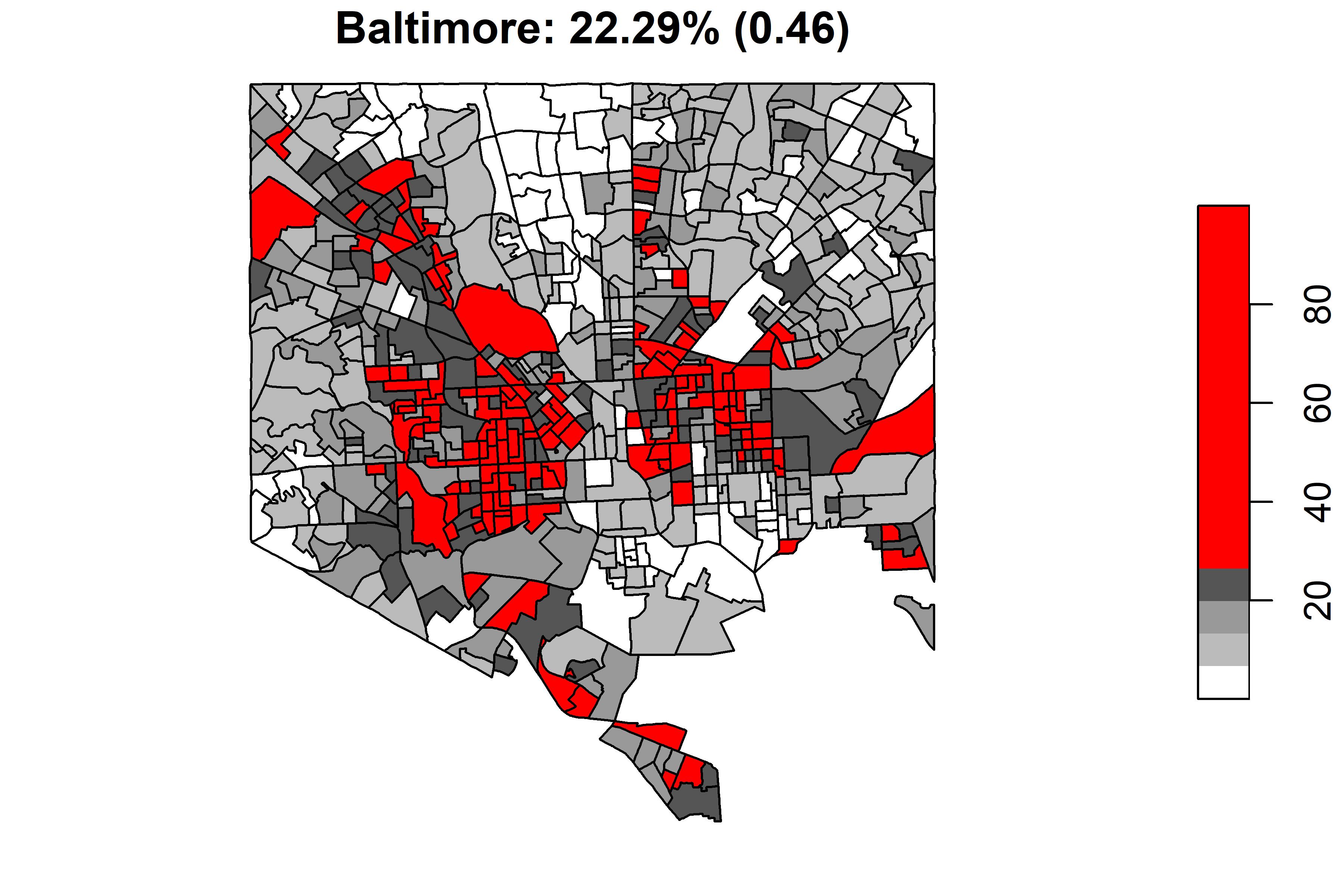}
\includegraphics[trim={1.5in 0 1in 0},clip,height=1.7in]{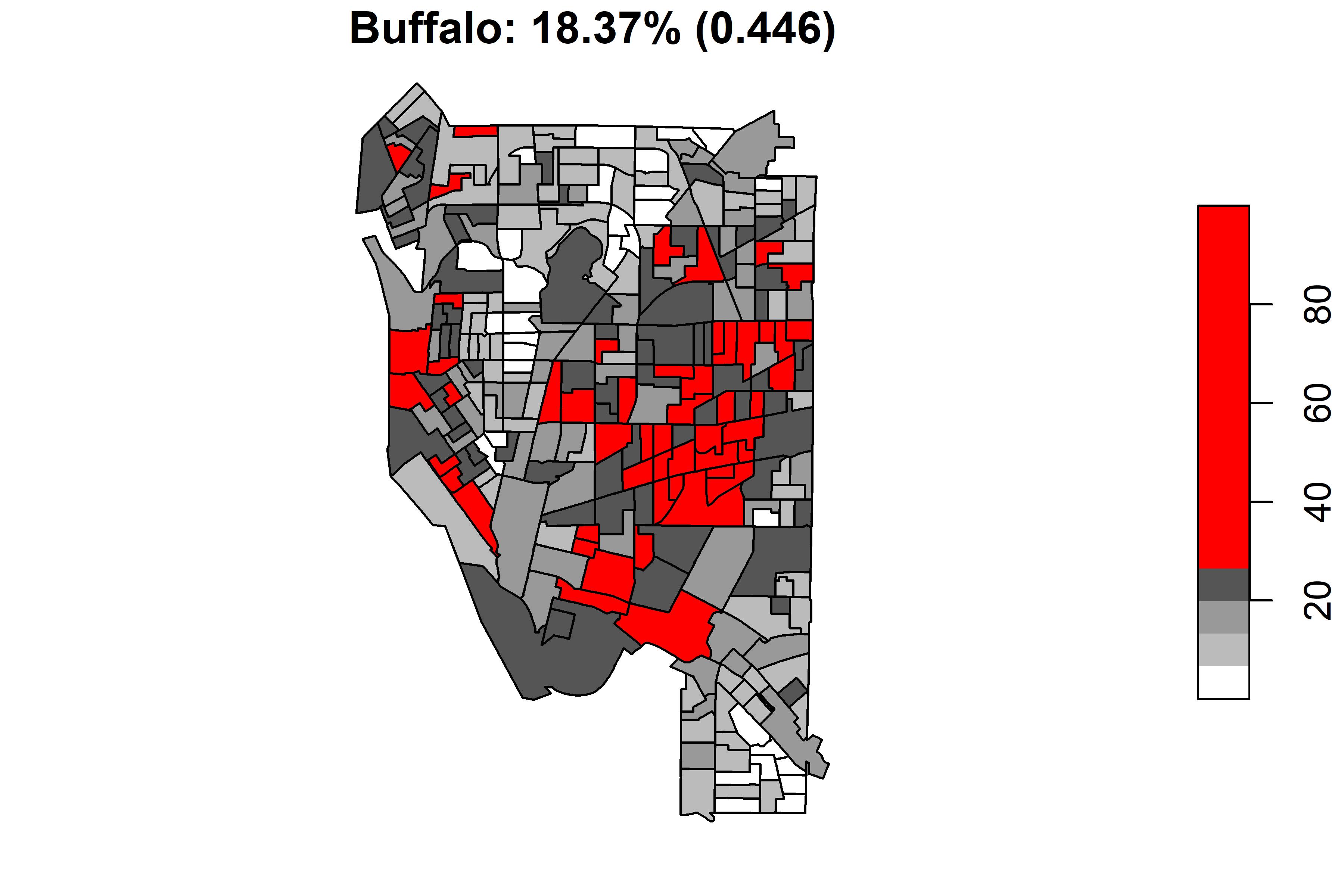}\\
\includegraphics[trim={1.5in 0 1.5in 0},clip,height=1.7in]{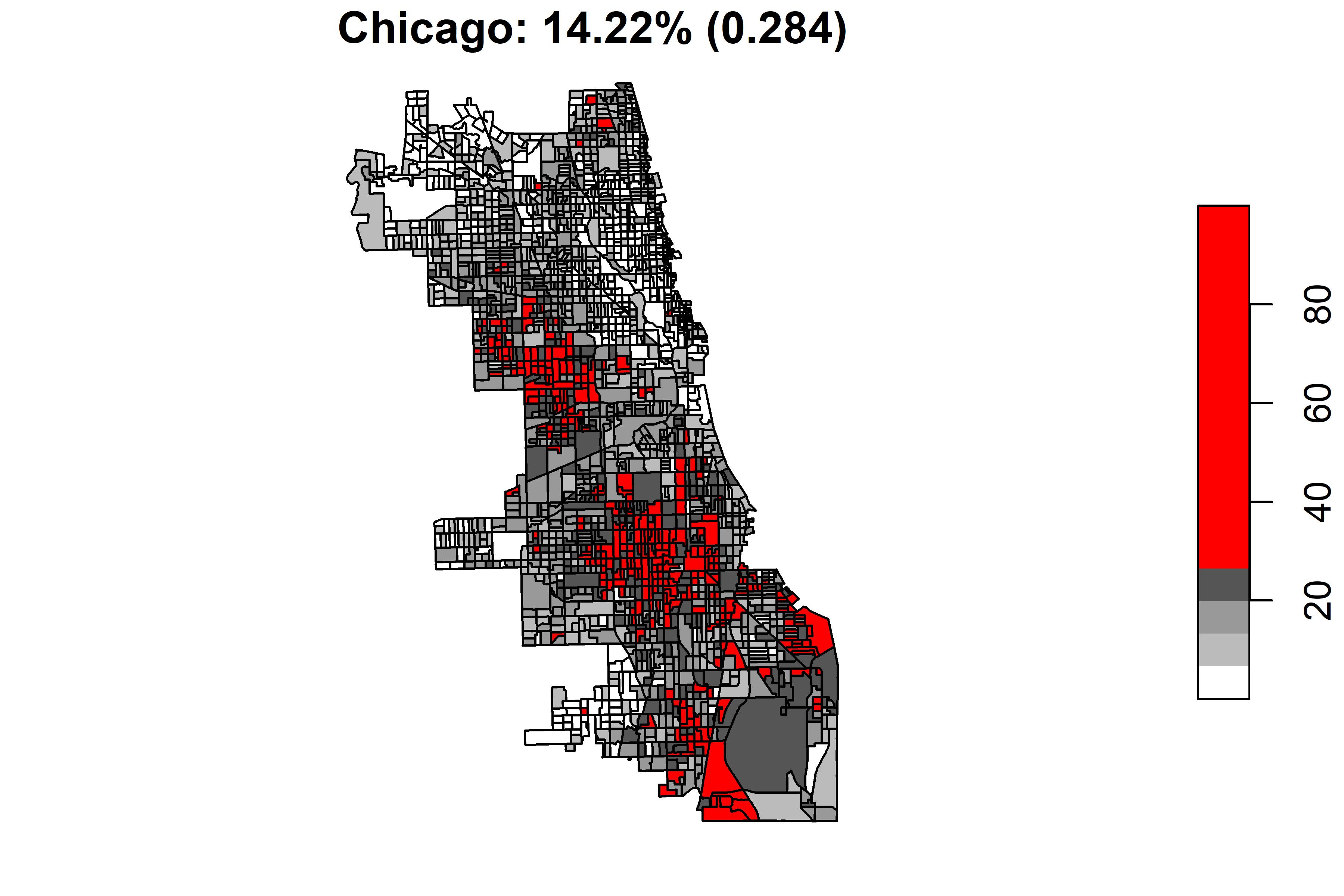}
\includegraphics[trim={.25in 0 1.5in 0},clip,height=1.7in]{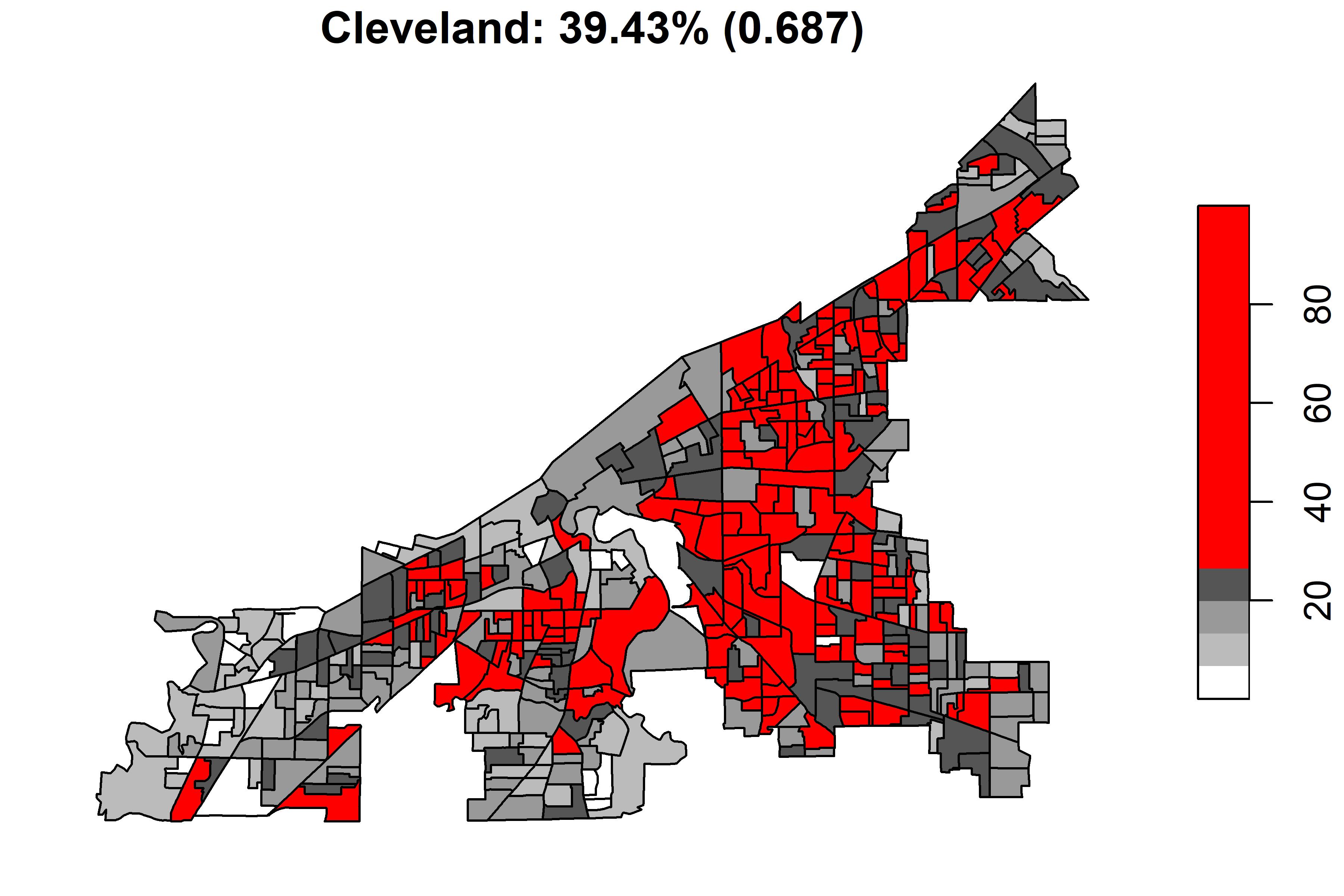}
\includegraphics[trim={.7in 0 1.1in 0},clip,height=1.7in]{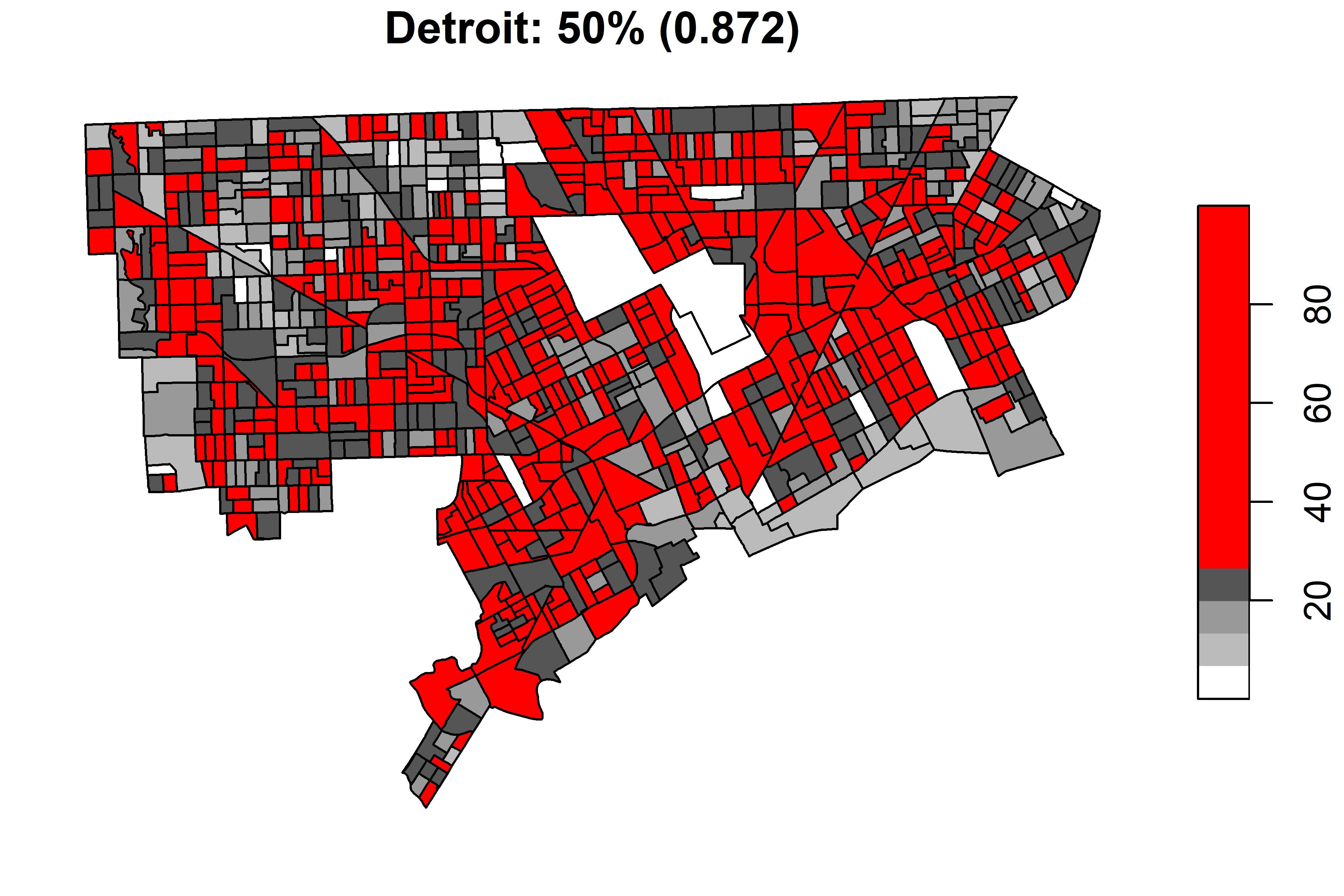}
\includegraphics[trim={.25in 0 1.5in 0},clip,height=1.7in]{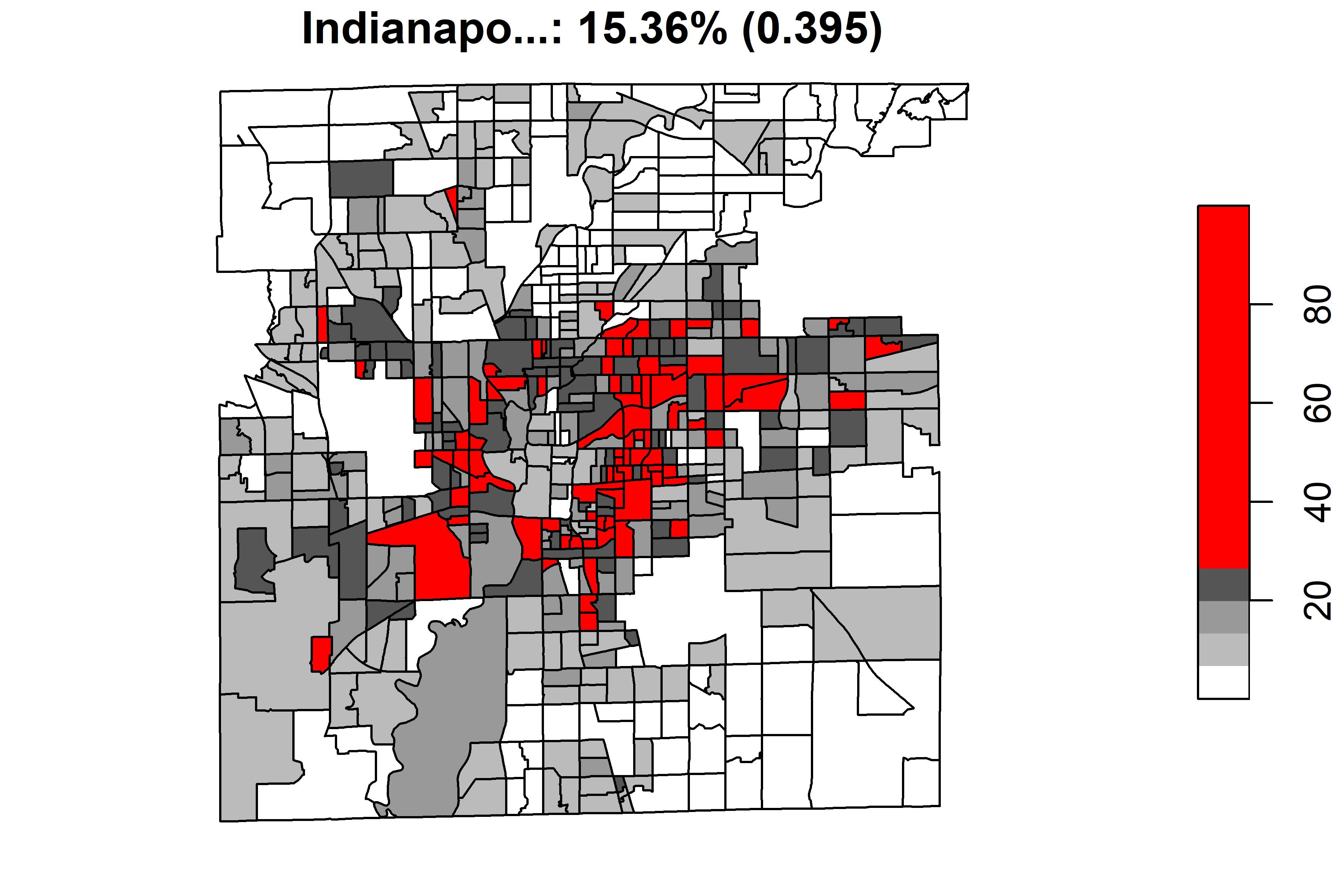}\\
\includegraphics[trim={.5in 0 1.2in 0},clip,height=1.7in]{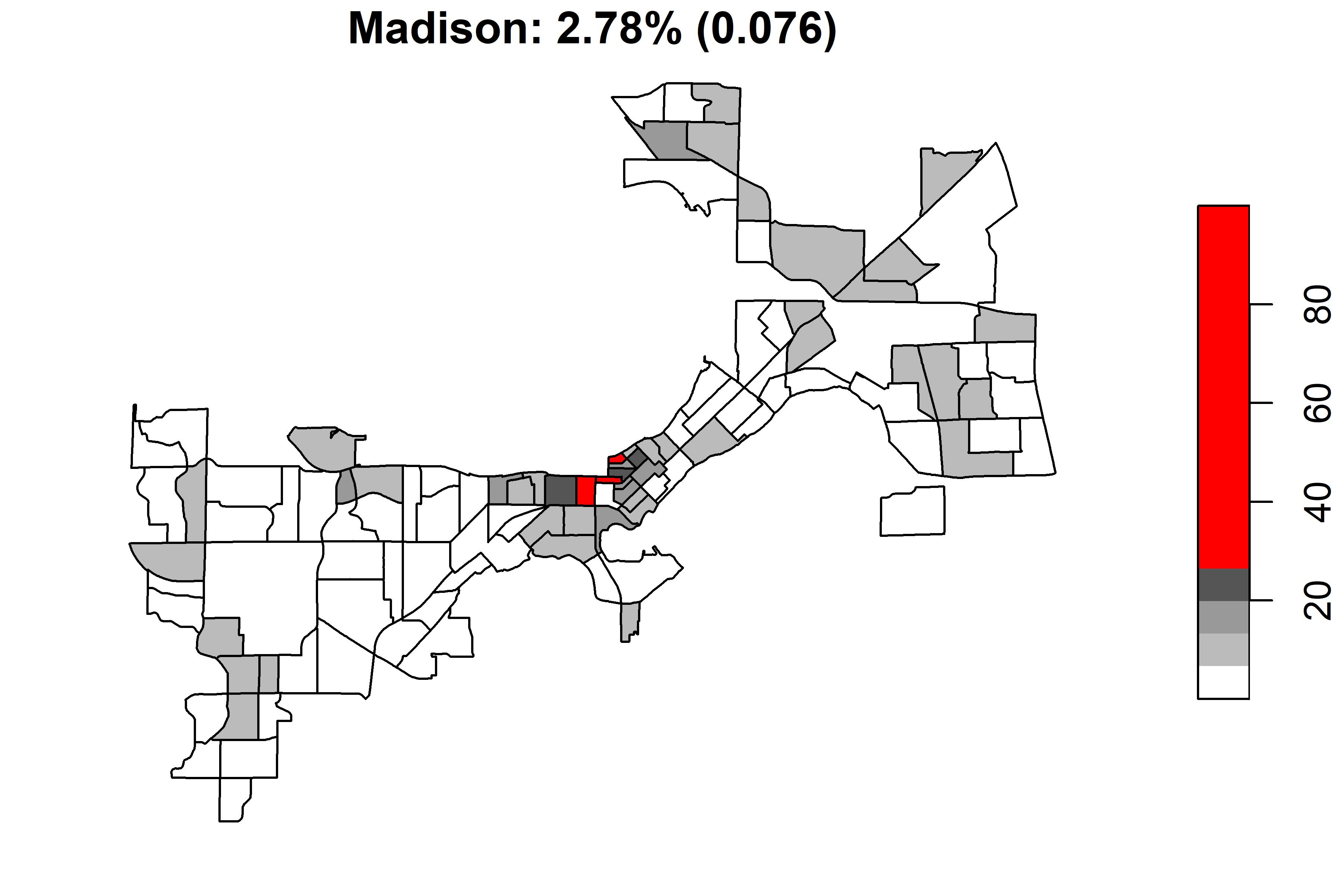}
\includegraphics[trim={1in 0 1.15in 0},clip,height=1.7in]{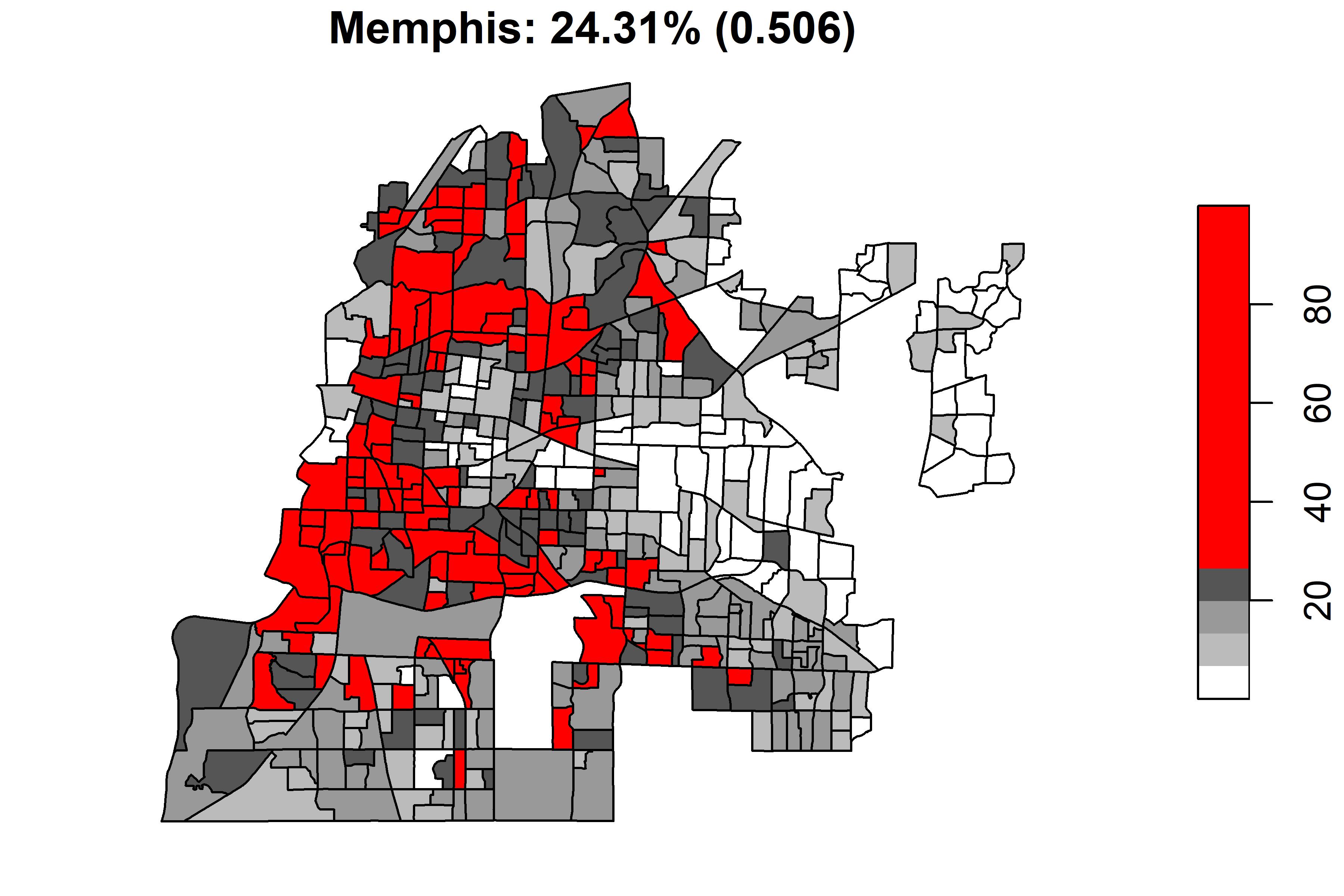}
\includegraphics[trim={.25in 0 1.5in 0},clip,height=1.7in]{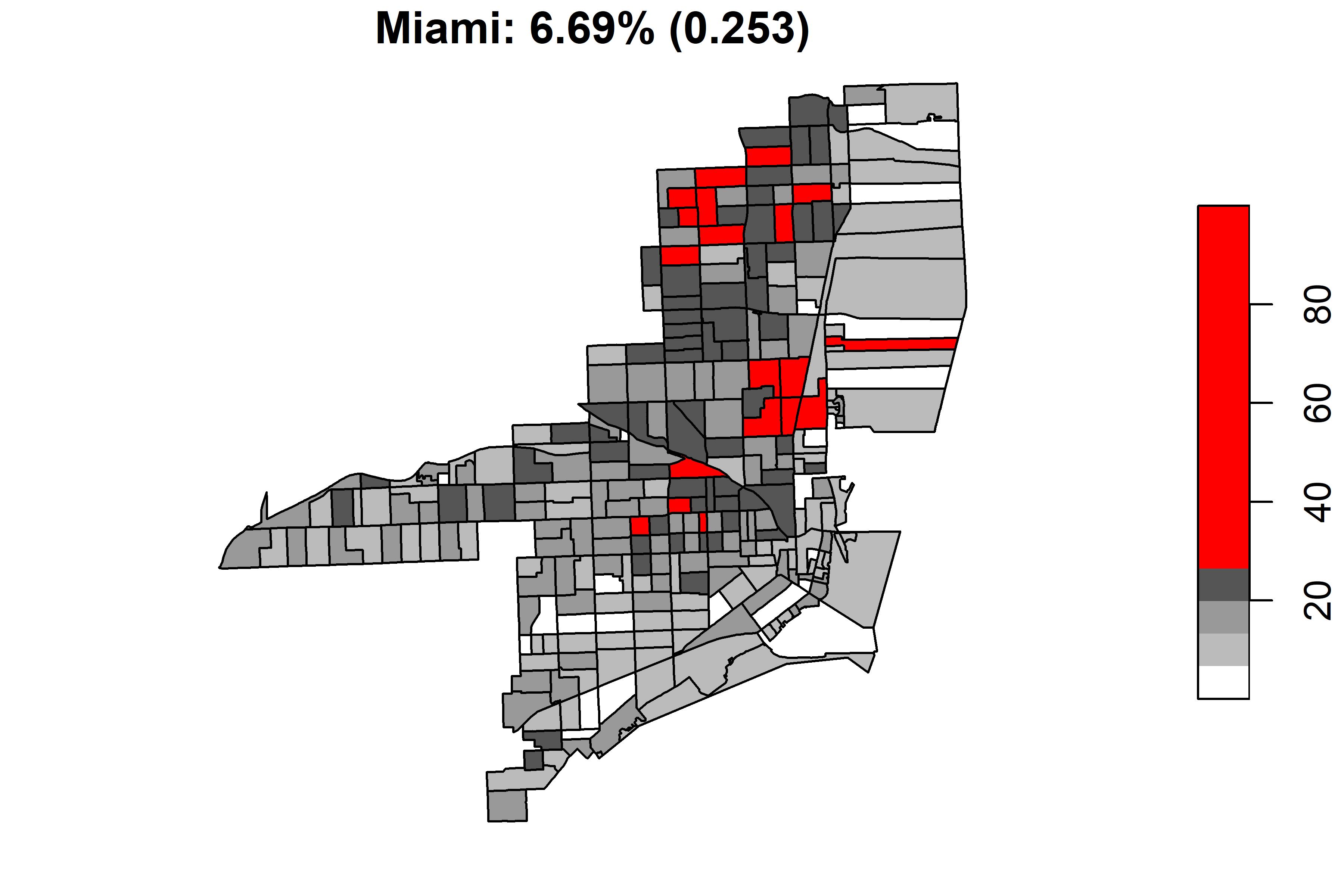}
\includegraphics[trim={1.25in 0 1.5in 0},clip,height=1.7in]{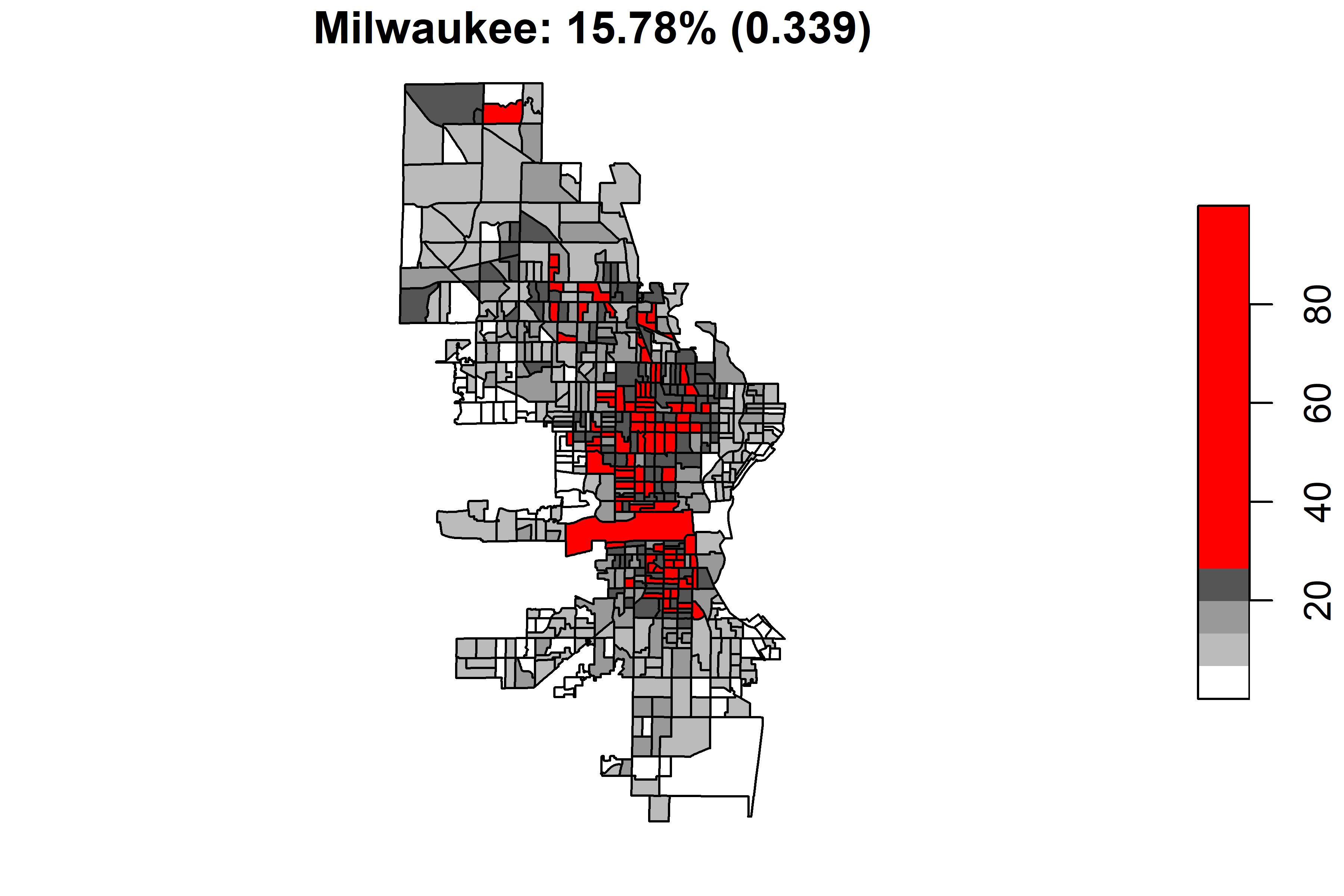}\\
\includegraphics[trim={1.25in 0 1.25in 0},clip,height=1.7in]{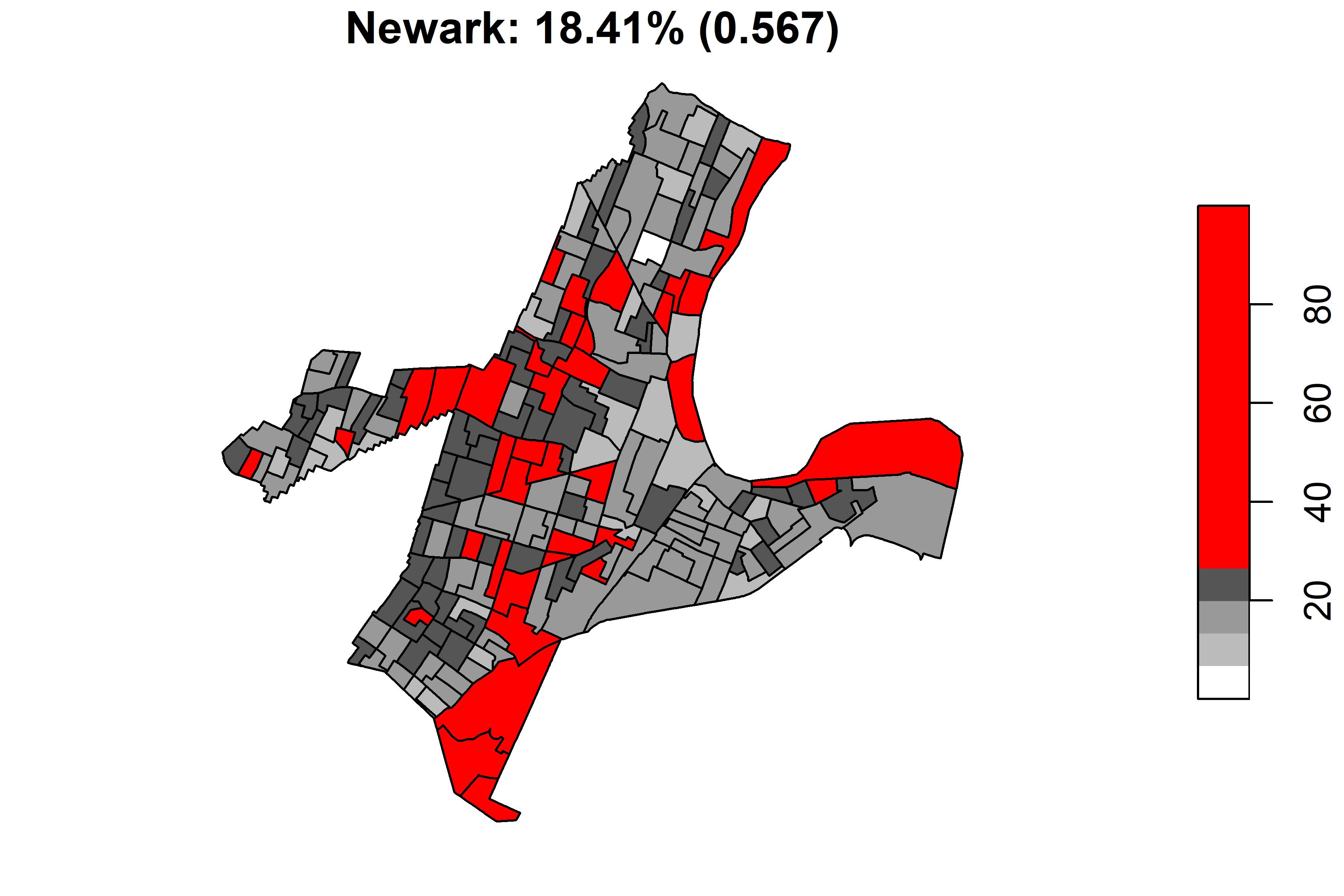}
\includegraphics[trim={1.25in 0 1.5in 0},clip,height=1.7in]{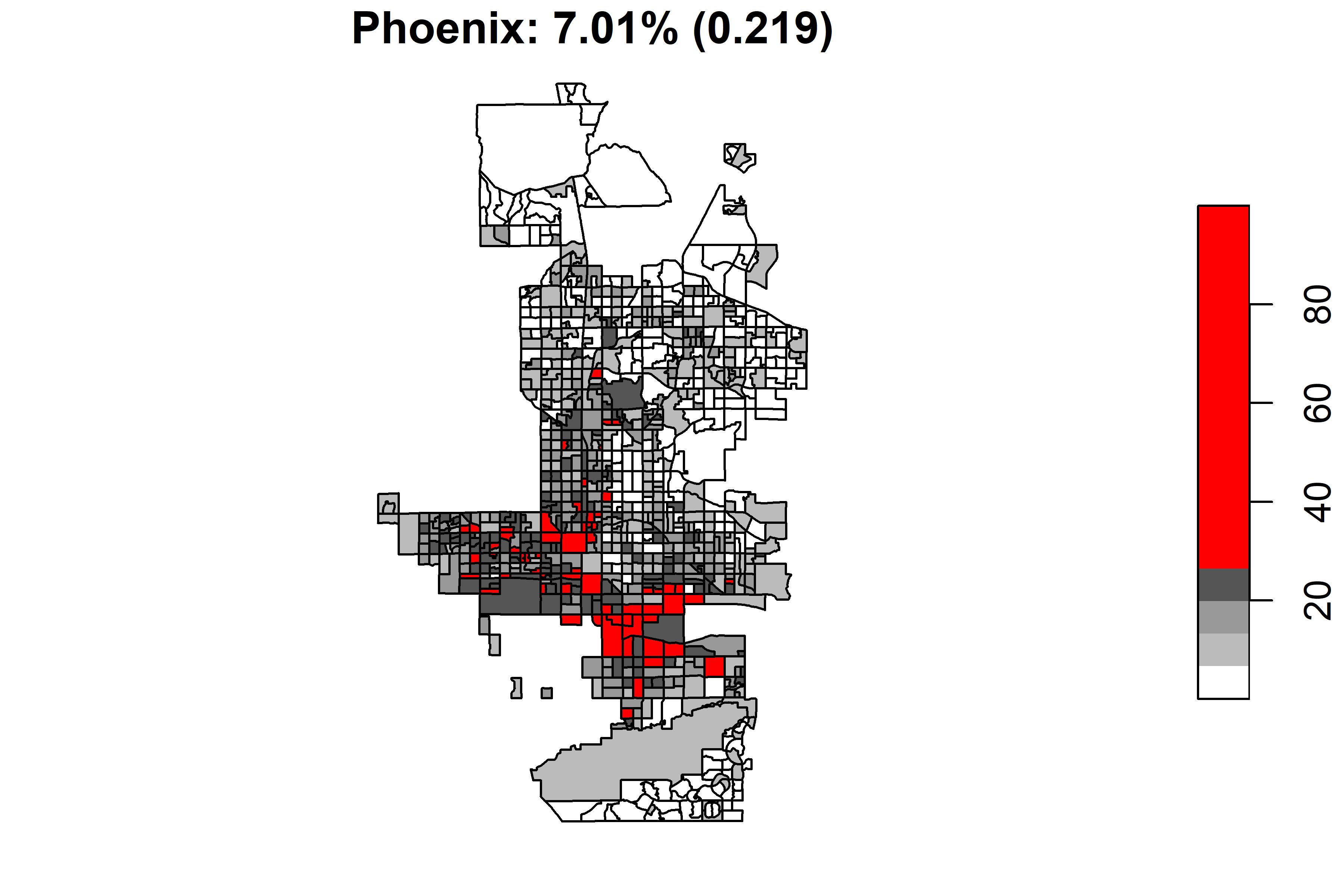}
\includegraphics[trim={.7in 0 1.5in 0},clip,height=1.7in]{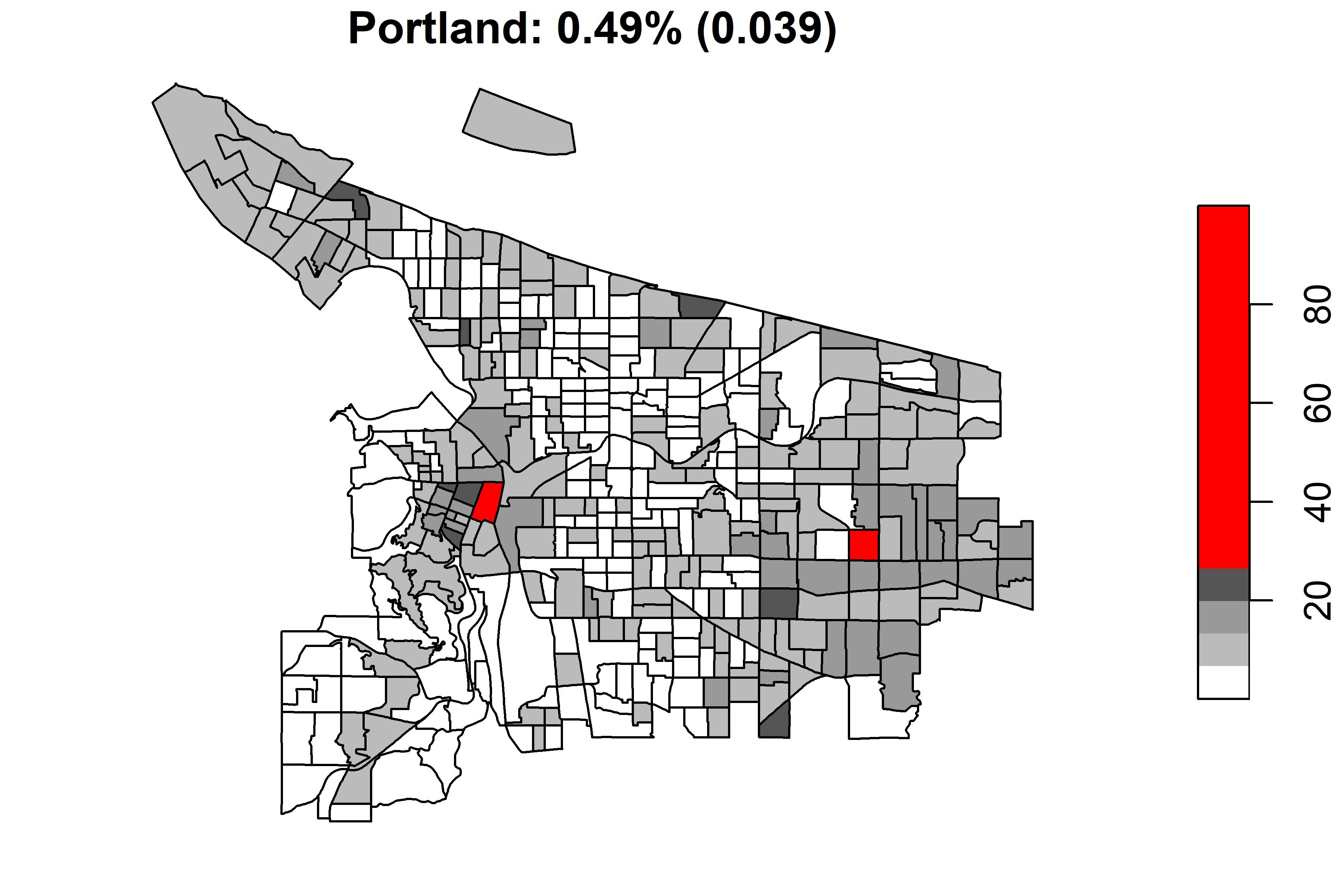}
\includegraphics[trim={1.25 0 1.25in 0},clip,height=1.7in]{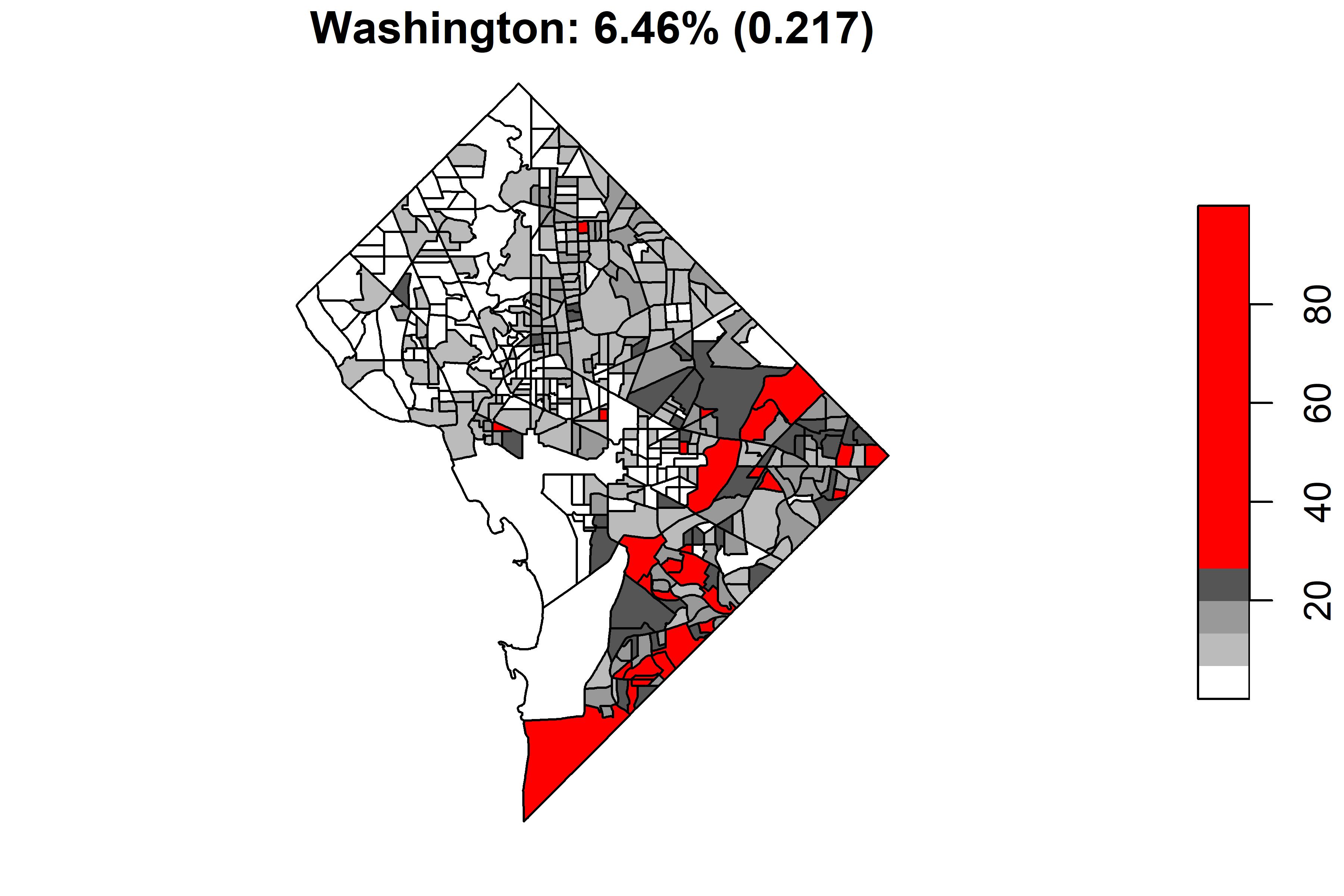}
\caption*{Percentage of high-deprivation areas and dispersion score (in parentheses).}
  \end{adjustwidth}
\end{figure}

\begin{table}[ht]
\begin{footnotesize}
\caption{Deprivation Scores, 83 Large Cities.}
  \begin{adjustwidth}{-1in}{-1in}  
        \begin{center}
\begin{tabular}{lrrrrlrrrr}
City&Code&Dep&\%HD&Disp&City&Code&Dep&\%HD&Disp\\
\hline
Albuquerque&ABQ&12.7&2.6&0.079&Louisville&LOU&10.8&1.7&0.069\\
Anaheim&ANA&16.2&1.9&0.105&Lubbock&LUB&10.2&4.4&0.157\\
Arlington&ARL&9.4&5.1&0.199&Madison&MAD&7.3&2.8&0.076\\
Atlanta&ATL&13.1&13.5&0.324&Memphis&MEM&19.6&24.3&0.506\\
Aurora&AUR&7.6&1.5&0.07&Mesa&MES&12.1&2.9&0.174\\
Austin&AUS&10.2&1.4&0.06&Miami&MIA&19.5&6.7&0.253\\
Bakersfield&BAK&14.4&17.9&0.148&Milwaukee&MKE&18.5&15.8&0.339\\
Baltimore&BAL&16.6&22.3&0.46&Minneapolis&MIN&9.1&3.5&0.186\\
Boston&BOS&14.3&6.2&0.26&Nashville&NAS&10.0&3&0.116\\
Buffalo&BUF&16.8&18.4&0.446&New Orleans&NLO&14.3&19.7&0.445\\
Chandler&CHN&5.8&0.8&0.05&New York&NEW&11.8&6&0.197\\
Charlotte&CHA&10.4&2.3&0.152&Newark&NEW&22.2&18.4&0.567\\
Chicago&CHI&15.4&14.2&0.284&Oakland&OAK&10.9&3.6&0.197\\
Chula Vista&CHV&10.6&7.5&0.337&Oklahoma City&OKC&12.9&11&0.291\\
Cincinnati&CIN&15.6&20.3&0.47&Omaha&OMA&10.1&4.7&0.177\\
Cleveland&CLE&25.6&39.4&0.687&Orlando&ORL&14.4&9&0.164\\
Colorado Spr...&COL&7.4&0&0&Philadelphia&PHL&14.3&16.4&0.394\\
Columbus&COL&10.6&16.5&0.387&Phoenix&PHX&14.5&7&0.219\\
Corpus Christi&CPC&14.0&8.4&0.263&Pittsburgh&PIT&11.1&6.5&0.183\\
Dallas&DAL&13.3&9&0.304&Plano&PLA&6.2&1.2&0.067\\
Denver&DEN&10.2&1.3&0.068&Portland&PDX&7.8&0.5&0.039\\
Detroit&DET&25.8&50&0.872&Raleigh&RAL&7.6&2.4&0.106\\
Durham&DUR&12.1&8.3&0.247&Riverside&RIV&14.5&4.6&0.163\\
El Paso&EL &12.4&12.8&0.33&Sacramento&SAC&12.9&4.4&0.178\\
Fort Wayne&FTW&10.3&10.7&0.293&San Antonio&SNT&14.3&7.8&0.284\\
Fort Worth&FWO&14.3&10.5&0.296&San Diego&SD&9.0&2.3&0.111\\
Fresno&FRE&15.1&27.6&0.494&San Francisco&SF&7.9&1.4&0.06\\
Greensboro&GRE&11.7&8.4&0.214&San Jose&SJ&10.8&1.1&0.058\\
Henderson&HEN&8.2&1.9&0.105&Santa Ana&STA&15.7&9.3&0.339\\
Houston&HOU&16.7&11.2&0.357&Scottsdale&SCO&10.7&0&0\\
Indianapolis&IND&11.5&15.4&0.395&Seattle&SEA&6.1&1.1&0.079\\
Irvine&IRV&6.6&0.9&0.045&St. Louis&STL&16.8&23.1&0.395\\
Jacksonville&JAX&9.8&9.4&0.236&St. Paul&STP&12.7&3.2&0.174\\
Jersey City&JER&10.1&2.4&0.113&St. Petersburg&SPT&14.4&6&0.196\\
Kansas City&KC&13.1&9.6&0.286&Stockton&STK&12.5&18.4&0.339\\
Laredo&LAR&15.2&26.6&0.47&Tampa&TAM&10.9&8.4&0.244\\
Las Vegas&LV&11.8&9.2&0.259&Toledo&TOL&14.4&25.5&0.41\\
Lexington&LEX&12.3&4.4&0.228&Tucson&TUC&13.0&7.8&0.295\\
Lincoln&LIN&6.4&0&0&Tulsa&TUL&15.4&5.8&0.161\\
Long Beach&LB&13.5&4.1&0.149&Virginia Beach&VB&7.0&0.3&0.02\\
Los Angeles&LAX&11.9&5.9&0.171&Washington&WAS&10.5&6.5&0.217\\
&&&&&Wichita&WIC&12.1&12&0.295\\
\hline

\end{tabular}
\end{center}
    \end{adjustwidth}
\caption*{Dep = Median deprivation score among all of a city's block groups.\\
Disp = Calculated dispersion score.}
\end{footnotesize}
\end{table}


\begin{table}[h]
\begin{footnotesize}
\caption{State-level Socioeconomic Characteristics.}
  \begin{adjustwidth}{-.5in}{-.5in}  
        \begin{center}
\begin{tabular}{lrrrrr}

State&Code&\%HD&PercPov&PercWhite&PercBlack\\
\hline
Mississippi&MS&14.1&20.3&58.4&37.7\\
Louisiana&LA&12.6&19.2&62.0&32.2\\
New Mexico&NM&10.5&19.1&74.8&2.1\\
Alabama&AL&10.1&16.7&68.1&26.6\\
Michigan&MI&9.6&14.4&78.4&13.8\\
Kentucky&KY&8&17.3&87.0&8.1\\
Ohio&OH&7.7&14.0&81.3&12.4\\
Arizona&AZ&7.3&15.1&77.2&4.5\\
Georgia&GA&7.2&15.1&58.6&31.6\\
West Virginia&WV&7.2&17.6&93.1&3.7\\
Texas&TX&7.1&14.7&74.0&12.1\\
Arkansas&AR&7&17.0&76.7&15.3\\
District of Columbia&DC&6.5&16.2&41.3&46.3\\
South Carolina&SC&6.4&15.2&67.2&26.8\\
Tennessee&TN&5.9&15.2&77.6&16.8\\
Nevada&NV&5.8&13.1&65.6&9.1\\
Illinois&IL&5.7&12.5&71.5&14.2\\
Indiana&IN&5.5&13.4&83.3&9.4\\
North Carolina&NC&5.4&14.7&68.7&21.4\\
Missouri&MO&5.3&13.7&82.2&11.5\\
New York&NY&4.9&14.1&63.7&15.7\\
California&CA&4.7&13.4&59.7&5.8\\
Oklahoma&OK&4.7&15.7&72.3&7.3\\
Pennsylvania&PA&4.7&12.4&80.5&11.2\\
Maryland&MD&4.6&9.2&55.5&29.9\\
Florida&FL&4.5&14.0&75.1&16.1\\
Connecticut&CT&4.1&9.9&75.9&10.7\\
Rhode Island&RI&4&12.4&80.5&6.8\\
South Dakota&SD&3.8&13.1&84.3&2.0\\
Kansas&KS&3.4&12.0&84.4&5.9\\
New Jersey&NJ&3.2&10.0&67.8&13.5\\
Wisconsin&WI&2.7&11.3&85.4&6.4\\
Delaware&DE&2.6&11.8&68.8&22.2\\
Massachusetts&MA&2.6&10.3&78.1&7.6\\
Virginia&VA&2.6&10.6&67.6&19.2\\
Maine&ME&2.5&11.8&94.3&1.4\\
Idaho&ID&1.6&13.1&90.0&0.7\\
Nebraska&NE&1.6&11.1&87.1&4.8\\
Montana&MT&1.5&13.1&88.5&0.5\\
Washington&WA&1.5&10.8&75.4&3.8\\
Iowa&IA&1.4&11.5&90.0&3.7\\
Oregon&OR&1.3&13.2&84.3&1.9\\
Utah&UT&1.2&9.8&86.4&1.2\\
Colorado&CO&1.1&10.3&84.0&4.2\\
Minnesota&MN&1&9.7&82.8&6.4\\
Vermont&VT&1&10.9&94.2&1.4\\
Wyoming&WY&1&11.0&91.4&1.0\\
North Dakota&ND&0.9&10.7&86.6&2.9\\
New Hampshire&NH&0.7&7.6&92.9&1.6\\
\hline

\end{tabular}
\end{center}
    \end{adjustwidth}
\caption*{}
\end{footnotesize}
\end{table}

\begin{table}[ht]
\begin{footnotesize}
\caption{Largest Places With High Deprivation Percentages.}
  \begin{adjustwidth}{-.5in}{-.5in}  
        \begin{center}
\begin{tabular}{lrrrrlrrrr}
Place $\ge$50\%&Pop&Place  = 100\%&Pop\\
\hline
Detroit city, Michigan&674841&Arvin city, California&21249\\
Flint city, Michigan&96559&Mendota city, California&11531\\
Calexico city, California&39946&Earlimart CDP, California&8668\\
Greenville city, Mississippi&30588&Orosi CDP, California&8300\\
Immokalee CDP, Florida&26597&Calipatria city, California&7395\\
Fort Hood CDP, Texas&23508&Huron city, California&7115\\
Hamtramck city, Michigan&21822&Newport city, Tennessee&6848\\
Prichard city, Alabama&21773&Mecca CDP, California&6635\\
Arvin city, California&21249&East Porterville CDP, California&6291\\
Kinston city, North Carolina&20398&Pahokee city, Florida&6269\\
Belle Glade city, Florida&19654&Monticello city, Kentucky&6070\\
Langley Park CDP, Maryland&19520&Warren city, Arkansas&5646\\
Boone town, North Carolina&19119&Palmview South CDP, Texas&5589\\
Milledgeville city, Georgia&18738&Brewton city, Alabama&5240\\
Imperial city, California&17454&Bret Harte CDP, California&5148\\
East Cleveland city, Ohio&17200&Keyser city, West Virginia&5050\\
Opelousas city, Louisiana&16234&Abbeville city, South Carolina&5045\\
North Myrtle Beach city, South Carolina&16200&Bolivar city, Tennessee&5031\\
Clearlake city, California&15349&Chinle CDP, Arizona&4879\\
Parlier city, California&15312&Ahoskie town, North Carolina&4848\\
\hline
\end{tabular}
\end{center}
    \end{adjustwidth}
\caption*{}
\end{footnotesize}
\end{table}

\begin{figure}[ht]
\hfill
\caption{National Map of Block-Group-Level Deprivation.}
  \begin{adjustwidth}{-1in}{-1in}  
\includegraphics[width=1.5\textwidth,angle=90]{NatlMap.jpg}
\caption*{Red = High-Deprivation.}
\end{adjustwidth}
\end{figure}

\nocite{Hegerty2019}
\nocite{Carstairs1995}
\nocite{townsend_1987}
\nocite{SMITH2009681}
\nocite{Pacione}
\nocite{Broadway}
\nocite{Langlois}
\nocite{Clelland}
\nocite{Burke}
\nocite{Noble}
\nocite{Kneebone}
\nocite{Morris1991}
\nocite{Salmond}
\nocite{Bertin}

\end{document}